\pdfoutput=1
\documentclass[aps,prd,amsmath,floats,floatfix, twocolumn,
superscriptaddress,showpacs,nofootinbib]{revtex4-2}

\usepackage[T1]{fontenc}
\usepackage[utf8]{inputenc}
\usepackage{lmodern}
\usepackage[caption=false]{subfig}
\usepackage{verbatim}
\usepackage{multirow}
\usepackage{mathrsfs}

\usepackage[dvipsnames, usenames]{xcolor}
\definecolor{linkcolor}{rgb}{0.0,0.3,0.5}
\usepackage[hypertexnames=false, unicode, colorlinks=true, linkcolor=linkcolor,
citecolor=linkcolor, filecolor=linkcolor,urlcolor=linkcolor,
pdfusetitle]{hyperref}

\usepackage[all]{hypcap}
\usepackage{graphicx}
\usepackage{xspace}
\usepackage{amssymb}
\usepackage[normalem]{ulem} %for \sout
\usepackage{bm} % boldmath
\usepackage{enumitem,amssymb}

\usepackage{microtype}
\usepackage{orcidlink}

\usepackage[english]{babel}
\usepackage{blindtext}

\graphicspath{%
  {figs/}%
}

\DeclareMathAlphabet{\mathpzc}{OT1}{pzc}{m}{it}

\newcommand{\etal}{\textit{et al.\ }}

\newlist{todolist}{itemize}{2}
\setlist[todolist]{label=$\square$}

\begin{document}

\title{Merging black holes with Cauchy-characteristic matching: Computing late-time tails}

\newcommand{\CornellLepp}{\affiliation{Laboratory for Elementary Particle Physics, Cornell University, Ithaca, New York 14853, USA}}
\newcommand{\CornellPhysics}{\affiliation{Department of Physics, Cornell University, Ithaca, NY, 14853, USA}}
\newcommand{\Cornell}{\affiliation{Cornell Center for Astrophysics
    and Planetary Science, Cornell University, Ithaca, New York 14853, USA}}
\newcommand\CornellPhys{\affiliation{Department of Physics, Cornell
    University, Ithaca, New York 14853, USA}}
\newcommand\Caltech{\affiliation{TAPIR 350-17, California Institute of
    Technology, 1200 E California Boulevard, Pasadena, CA 91125, USA}}
\newcommand{\AEI}{\affiliation{Max Planck Institute for Gravitational Physics
    (Albert Einstein Institute), Am M\"uhlenberg 1, Potsdam 14476, Germany}} %
\newcommand{\UMassD}{\affiliation{Department of Mathematics,
    Center for Scientific Computing and Visualization Research,
    University of Massachusetts, Dartmouth, MA 02747, USA}}
\newcommand\Olemiss{\affiliation{Department of Physics and Astronomy,
    The University of Mississippi, University, MS 38677, USA}}
\newcommand{\Bham}{\affiliation{School of Physics and Astronomy and Institute
    for Gravitational Wave Astronomy, University of Birmingham, Birmingham, B15
    2TT, UK}}
\newcommand{\Perimeter}{\affiliation{Perimeter Institute for Theoretical Physics, Waterloo, ON N2L2Y5, Canada}}

\author{Sizheng Ma \orcidlink{0000-0002-4645-453X}}
\email{sma2@perimeterinstitute.ca}
\Perimeter

\author{Mark A. Scheel \orcidlink{0000-0001-6656-9134}}
\Caltech

\author{Jordan Moxon \orcidlink{0000-0001-9891-8677}}
\Caltech

\author{Kyle C. Nelli \orcidlink{0000-0003-2426-8768}}
\Caltech

\author{Nils Deppe \orcidlink{0000-0003-4557-4115}}
\CornellLepp
\CornellPhys
\Cornell

\author{Lawrence E.~Kidder\orcidlink{0000-0001-5392-7342}}
\Cornell

\author{William Throwe \orcidlink{0000-0001-5059-4378}}
\Cornell

\author{Nils L.~Vu \orcidlink{0000-0002-5767-3949}}
\Caltech

% Because hyperref only gets the *last* author, we need to be explicit.
\hypersetup{pdfauthor={Ma et al.}}

\date{\today}

%==========================================================================
\begin{abstract}
Cauchy-characteristic matching (CCM) is a numerical-relativity technique that solves Einstein's equations on an effectively infinite computational domain, thereby eliminating systematic errors associated with artificial boundary conditions.
Whether CCM can robustly handle fully nonlinear, dynamical spacetimes, such as binary black hole (BBH) mergers, has remained an open question.
In this work, we provide a positive answer by presenting nine successful CCM simulations of BBHs; and demonstrate a key application of this method: computing late-time tails.
Our results pave the path for systematic studies of late-time tails in BBH systems, and for producing highly accurate waveforms essential to next-generation gravitational-wave detectors.
\end{abstract}

\maketitle

% \textcolor{PineGreen}{
% \begin{todolist}
%    % \item[\checkmark] Mention the analytical exploration of likelihood (marginalization).
%     \item mention priors on mass and spin etc.
%     % \item traditional $\to$ full-RD
% \end{todolist}
% }
%\done
%==========================================================================
\section{Introduction}
\label{sec:introduction}
The detection of gravitational waves (GWs) by the LIGO, Virgo, and KAGRA collaboration \cite{TheLIGOScientific:2016pea, LIGOScientific:2018mvr, LIGOScientific:2020ibl, LIGOScientific:2021djp, TheLIGOScientific:2014jea,TheVirgo:2014hva,Somiya:2011np} has revolutionized our ability to explore the strong-field regime of gravity in compact binary systems. 
A key ingredient in GW astronomy is the accurate modeling of GW signals from binary black hole (BBH) mergers --- the dominant sources in current observations --- to reliably extract them from detector noise.

To date, numerical relativity (NR) remains the only \textit{ab initio} method capable of computing the full nonlinear waveforms from BBHs. In particular,
NR simulations are used to calibrate effective-one-body models \cite{Pan:2009wj,Bohe:2016gbl,Pompili:2023tna,Nagar:2018zoe,Nagar:2020pcj} and to construct surrogate waveforms \cite{Blackman:2017pcm,Varma:2018mmi,Varma:2019csw,Yoo:2023spi}, both of which have become essential tools in LIGO-Virgo-KAGRA data analysis \cite{TheLIGOScientific:2016pea, LIGOScientific:2018mvr, LIGOScientific:2021djp,Abbott:2020gyp,LIGOScientific:2020tif,LIGOScientific:2020ibl,LIGOScientific:2020tif,Abbott:2020tfl,LIGOScientific:2020stg,Abbott:2020mjq,LIGOScientific:2021tfm,LIGOScientific:2021izm}. However, meeting the stringent requirements of next-generation GW detectors will require NR accuracy to improve by an order of magnitude \cite{Purrer:2019jcp}.
Systematic errors in NR waveforms --- if left unaddressed --- could lead to false alarms of deviations from general relativity. Therefore, reducing or eliminating these systematics is a critical priority for future GW astronomy.

The most commonly used approach in NR is the Cauchy formulation, in which spacetime is foliated into spacelike hypersurfaces and Einstein's equations are decomposed into evolution and constraint sets via the 3+1 decomposition (see, e.g., \cite{Baumgarte_Shaprio:NumRel}). Since the Cauchy formulation is restricted to a finite computational domain, an artificial outer boundary must be introduced, along with appropriate boundary conditions. These conditions typically approximate, rather than reproduce exactly, the behavior of gravitational fields that extend to infinity and can thus become a significant source of systematic error, e.g., see \cite{Allen:2004js, Dafermos:2004wt}. To overcome this issue, it is desirable to develop NR frameworks that directly incorporate null infinity, where GWs are unambiguously defined.

One promising method is to use hyperboloidal slices \cite{Zenginoglu:2008pw,Rinne:2013qc,Vano-Vinuales:2014koa,Vano-Vinuales:2014ada,Morales:2016rgt,Vano-Vinuales:2024tat,Peterson:2024bxk}, which are spacelike hypersurfaces that approach future null infinity $\mathscr{I}^+$. By introducing a compactified radial coordinate, future null infinity is brought to a finite coordinate distance on the 
computational domain, thereby removing artificial outer boundaries and allowing GWs to be extracted directly at $\mathscr{I}^+$.
Another strategy is based on the generalized conformal field equations \cite{friedrich1998gravitational,Frauendiener:2022bkj,Frauendiener:2023ltp,Frauendiener:2021eyv,Frauendiener:2025xcj}, which simultaneously include both past and future null infinities $\mathscr{I}^\pm$ within a unified computational framework.
This enables the study of the global structure of spacetime and scattering processes between $\mathscr{I}^-$ and $\mathscr{I}^+$ \cite{Frauendiener:2025xcj}. However, current numerical implementations of both approaches are largely limited to single BHs or simple spacetimes (with symmetries), and their extension to fully generic BBH mergers remains a challenge.

A third option to include null infinity is the characteristic formulation \cite{winicour2012characteristic,Bishop_1993,Bishop:1996gt,Bishop:1997ik,Bishop:1998uk,Barkett:2019uae,Moxon:2020gha,Moxon:2021gbv}, in which spacetime is foliated by null hypersurfaces that extend to $\mathscr{I}$. As in the hyperboloidal approach, a compactified radial coordinate maps null infinity to the finite boundary of the computational domain, which in turn enables unambiguous extraction of GWs at $\mathscr{I}$.
However, the characteristic method faces challenges in evolving the near-zone region of BBH spacetimes due to the formation of caustics --- regions where neighboring null rays focus and intersect --- resulting in coordinate singularities
\cite{corkill1983numerical,friedrich1983characteristic,Stewart1986,Frittelli:1997rw,Bhagwat:2017tkm,Baumgarte:2023tdh}. Such singularities undermine the stability and accuracy of the characteristic simulation in highly dynamical or strongly curved regions.

Since the Cauchy and characteristic formulations are naturally complementary, they can be combined to achieve a complete description of gravitational fields. In this hybrid approach, the Cauchy system first evolves the near-zone dynamics and provides metric data on a timelike worldtube, which is then used as input for the characteristic evolution that propagates the data to future null infinity. This two-step method for extracting gravitational radiation is known as Cauchy-characteristic evolution (CCE) \cite{winicour2012characteristic,doi:10.1063/1.525904,1992anr..conf...20B,1983JMP....24.1193W,doi:10.1063/1.526472,1997JCoPh.136..140B,Bishop:1996gt,Bishop:1997ik,Bishop:1998uk,Szilagyi:2000xu,Barkett:2019uae,Moxon:2020gha,Moxon:2021gbv,gomez1994null,Bishop:1996gt,Bishop:1997ik,Gomez:2001pb,Bishop:2003bz,Reisswig:2006nt,Gomez:2007cj,Gomez:2007cj,Babiuc:2008qy,Reisswig:2012ka,Reisswig:2009us,Reisswig:2009rx,Babiuc:2010ze,Babiuc:2011qi,Handmer:2014qha,Handmer:2015dsa,Handmer:2016mls,Barkett:2019uae,Moxon:2020gha,Moxon:2021gbv,Ma:2024bed,Mitman:2020pbt,Mitman:2024uss}. The first CCE simulation of BBH systems was given by Reisswig \etal \cite{Reisswig:2009us}, and the method has been applied to studies of memory effects \cite{Mitman:2020pbt,Mitman:2024uss}. Despite recent fruitful progress, CCE has a notable limitation: its data flow is strictly one-way. The characteristic evolution depends on the Cauchy data, but not vice versa --- meaning the Cauchy simulation must still be run with artificial outer boundary conditions. As a result, while CCE offers a robust and gauge-invariant method for extracting waveforms at $\mathscr{I}^+$, it does not eliminate systematic errors introduced by inaccurate Cauchy boundary conditions.

\textit{Cauchy-characteristic matching} (CCM) \cite{Bishop_1993,PhysRevLett.76.4303,winicour2012characteristic} is a natural solution to this limitation. Unlike CCE, where the Cauchy and characteristic evolutions are performed sequentially, CCM evolves both systems simultaneously, with each providing \emph{exact} boundary data for the other. The Cauchy evolution supplies interior data to the characteristic domain at the worldtube, while the characteristic system, in turn, provides outer boundary data for the Cauchy evolution. This mutual exchange eliminates the need for artificial boundary conditions entirely and allows for a seamless and self-consistent evolution from the strong-field near zone out to future null infinity.

Despite the conceptual appeal of CCM, its numerical implementation has proven challenging \cite{winicour2012characteristic}. Early efforts focused on spacetimes with spherical symmetry \cite{PhysRevD.58.044019,Gomez:1996dy}, cylindrical symmetry \cite{PhysRevD.52.6863,PhysRevD.52.6868,dInverno:2000uvh}, and axial symmetry \cite{PhysRevD.54.4919,PhysRevD.56.772}. In three dimensions, CCM has only been implemented perturbatively \cite{BinaryBlackHoleGrandChallengeAlliance:1997aaw,Rupright:1998uw,Rezzolla:1998xx}: the wave zone was not evolved fully nonlinearly with a characteristic code, but instead modeled using linearized perturbations of Schwarzschild spacetime. This approach yielded simulations of 3D Teukolsky waves propagating on a flat background \cite{BinaryBlackHoleGrandChallengeAlliance:1997aaw}, but left unresolved the feasibility of applying CCM to fully nonlinear, dynamical spacetimes like BBH mergers. More critically, it was later pointed out that CCM may be only weakly hyperbolic \cite{winicour2012characteristic,Giannakopoulos:2023nrb,Giannakopoulos:2020dih,Giannakopoulos:2021pnh,Giannakopoulos:2023zzm,Gundlach:2024xmo}, implying that a stable numerical implementation may not exist in general. This concern further cast doubt on the long-term viability of the CCM program\footnote{See Sec.~\ref{sec:conclusion} for more discussions.}.

Recently, we developed a fully relativistic algorithm to perform CCM for \emph{any} numerical spacetime \cite{Ma:2023qjn} (hereafter Paper I). The method is free of approximations and has already demonstrated stability in simulations of simple systems such as single BHs and flat spacetime. In this paper, we apply the same algorithm to BBH mergers and show that CCM is not only feasible in this highly dynamical regime but also yields stable, convergent, and high-accuracy simulations. As discussed in Paper I, a key advantage of CCM is its ability to model backscattering effects, which are closely related to GW tails \cite{PhysRevD.5.2419,PhysRevD.34.384,1975ApJ...200..245T,RevModPhys.52.299,PhysRevD.37.1410}. In what follows, we demonstrate that our CCM framework can indeed resolve the late-time tails arising from BBH mergers.

This paper is organized as follows. In Sec.~\ref{sec:summary}, we briefly summarize the CCM algorithm designed in Paper I. Next in Sec.~\ref{sec:numerical_simulations}, we present nine CCM simulations of BBH mergers, along with late-time tails uncovered through this method. These tails are then analyzed phenomenologically in Sec.~\ref{sec:tail_analysis}. Finally, we summarize the results in Sec.~\ref{sec:conclusion}.

Throughout this paper we use Latin indices $i, j, k, \ldots$ to denote 3D
spatial components and Greek indices $\mu,\nu,\ldots$ for 4D spacetime
components. The initial ADM mass of binary systems is denoted by $M$.

%==========================================================================
\section{Summary of CCM}
\label{sec:summary}
In this section, we provide a brief overview of the CCM algorithm developed in Paper I and refer the reader to the paper for more details.

\begin{table*}
    \centering
    \caption{List of BBH simulations considered in this work. Here $q\geq 1$ stands for the mass ratio. Each binary is initialized with an orbital angular velocity $\Omega_{\rm orb}$ (the transverse velocity relative to the line connecting the two BHs) and a radial velocity $\dot{r}$. The merger time corresponds to the formation of a common apparent horizon. The remnant Kerr BH has mass $M_f$ and dimensionless spin $\chi_f$. CPU hours are provided for each simulation. Reference systems involve sequential Cauchy and characteristic evolutions, their CPU times are listed in the format (Cauchy)+(characteristic).}
    \begin{tabular}{c c c c c c c c c c c} \hline\hline
   \multirow{2}{*}{Systems} & CCM or & \multirow{2}{*}{$q$} & $M\Omega_{\rm orb}$ & $\dot{r}$ & Initial Separation  & Boundary Radius & Merger Time &  \multirow{2}{*}{$M_f/M$}   & \multirow{2}{*}{$\chi_f$}    & CPU Time   \\ 
& Reference & & $(10^{-5})$  & $(10^{-4})$ & $(M)$ & $(M)$  & $(M)$ & & & (h)  \\ \hline
 \multirow{6}{*}{Head-on } & CCM & 1 &  0 & 0 & 110 & 650 & 1323.7 & 0.99712 & $1.1\times 10^{-9}$ & 3263.4 \\ \cline{2-11} 
 & Reference  & 1 &  0 & 0 & 110 & 6000 & 1323.7 & 0.99712 & $4.0\times 10^{-10}$ & 10211.2+2\\ \cline{2-11}
 & CCM  & 2 &  0 & 0 & 110 & 650 & 1323.8 & 0.99752 & $3.7\times10^{-9}$ & 5556.7 \\ \cline{2-11}
 & Reference  & 2 &  0 & 0 & 110 & 4672 & 1323.8 & 0.99752 & $2.5\times10^{-9}$ & 11297.1+2 \\ \cline{2-11}
 & CCM   & 4 &  0 & 0 & 110 & 650 & 1323.7 & 0.99830 & $2.1\times10^{-9}$ & 10289.1 \\ \cline{2-11}
 & Reference & 4 &  0 & 0 & 110 & 4800 & 1323.8 & 0.99830 & $2.2\times10^{-9}$ & 18770.3+2   \\ \hline
  \multirow{8}{*}{\shortstack{Quasi \\Head-on }} & CCM & 1 & 8.562 & 0 & 110 & 650 & 1332.9 & 0.99655 & 0.259 & 3128.1 \\ \cline{2-11}
  & Reference & 1 & 8.562 & 0 & 110 & 4800  & 1332.9 & 0.99655 & 0.259 & 8974.3+2  \\ \cline{2-11}
  & CCM  & 1  & 17.123 & 0 & 110 & 650 & 1362.3 & 0.99340 & 0.509 & 3476.1 \\ \cline{2-11}
  & Reference  & 1  & 17.123  & 0 & 110 & 4669 & 1362.3 & 0.99340 & 0.509  & 8766.5+2 \\ \cline{2-11}
  & CCM & 1  & 25.686 & 0 & 110 & 650 & 1422.2 & 0.97602 & 0.715 & 5022.7\\ \cline{2-11}
  & Reference & 1  & 25.686 & 0 & 110 & 4800  & 1422.2 & 0.97602 & 0.715 & 14641.8+2  \\ \cline{2-11}
  & CCM & 1  & 29.967 & 0 & 110 & 650 & 1498.8 & 0.94480 & 0.696 & 6775.1 \\ \cline{2-11}
  & Reference & 1  & 29.967  & 0 & 110 & 4800  & 1498.8 & 0.94480 & 0.696 & 17556.9+2  \\ \hline
  Eccentric & CCM & 1 & 58.102 & 0.31 & 90 & 650 & 14781 & 0.95272 & 0.689 & 14318.6  \\ \hline
  \multirow{2}{*}{\shortstack{Quasi \\Circular }}  &\multirow{2}{*}{CCM} &\multirow{2}{*}{1} &\multirow{2}{*}{1979.005}  &\multirow{2}{*}{2.6} & \multirow{2}{*}{13} & \multirow{2}{*}{300} & \multirow{2}{*}{2561.8} & \multirow{2}{*}{0.95159} & \multirow{2}{*}{0.686} & \multirow{2}{*}{3178.1}\\ 
  \\
  \hline\hline
     \end{tabular}
     \label{table:NR_runs}
\end{table*}

Our Cauchy evolution adopts the Generalized Harmonic (GH) formalism described in
\cite{Lindblom:2005qh}, which formulates the vacuum Einstein equations as first-order symmetric hyperbolic partial differential equations. This
system evolves fifty variables: the metric
$g_{\mu\nu}$, its normal-time derivative
$\Pi_{\mu\nu}=\alpha^{-1}(\beta^i\partial_ig_{\mu\nu}-\partial_tg_{\mu\nu})$,
and spatial derivative $\Phi_{i\mu\nu}=\partial_ig_{\mu\nu}$,
where $\alpha$ and $\beta^i$ are the lapse and shift, respectively.

At the outer boundary of the Cauchy domain, boundary conditions are imposed on forty incoming
characteristic fields, including \cite{Kidder:2004rw,Lindblom:2005qh}
\begin{equation}
\begin{aligned}
    &u^{0}_{\mu\nu}=g_{\mu\nu}, \quad u^{2}_{i\mu\nu}=P_{i}^k\Phi_{k\mu\nu},  \\
    &u^{1-}_{\mu\nu}=\Pi_{\mu\nu}-s^i\Phi_{i\mu\nu}-\gamma_2\psi_{\mu\nu},
\end{aligned}
\end{equation}
where $s^i$ is the outward unit normal vector of the boundary,
$P_{i}^k=\delta_{i}^k-s_is^k$ is the projection operator, and $\gamma_2$ is a
constraint damping parameter. As noted in \cite{Lindblom:2005qh}, the boundary conditions can be divided into three subsets: constraint, physical, and gauge.

In the constraint subset, the normal derivatives of $u^{0}_{\mu\nu}$,
$u^{2}_{i\mu\nu}$, and four components of $u^{1-}_{\mu\nu}$ are related to
incoming constraint modes 
[see Eq.~(59)-(61) in \cite{Lindblom:2005qh}]. Imposing constraint-preserving conditions yields
Neumann conditions on these thirty-four variables. Thus, CCM is unnecessary here, as manual constraint injection lacks physical motivation and may cause
numerical instability. 

The physical subset involves two other components of $u_{\mu\nu}^{1-}$ and encodes information about backscattered GWs. According to Eq.~(66) and (67) in
\cite{Lindblom:2005qh} and Eq. (2.12) in Paper I, these degrees of freedom are related to the Weyl scalar $\Psi_0$, corresponding to the two polarizations of the backscattered radiation. Setting boundary conditions for this subset requires precisely modeling the
backscattered waves outside the Cauchy domain, which was achieved in
Paper I. Since the two components of $u_{\mu\nu}^{1-}$ are represented on the
Cauchy grid with the Cauchy tetrad, careful gauge and tetrad transformations are
necessary before matching. We have verified that neglecting these transformations can cause instability in CCM simulations. In Paper I, we developed a general method for
handling these transformations in any numerical spacetime.

In the wave zone, a numerical spacetime can be approximated as a perturbed
Schwarzschild BH, allowing the use of the Regge-Wheeler-Zerilli (RWZ) formalism
to estimate the evolution of $\Psi_0$. This
approach leads to “higher order boundary conditions”
(HOBCs)~\cite{Buchman:2006xf, Buchman:2007pj, Rinne:2008vn, Buchman:2024zsb}
that absorb
outgoing multipolar radiation up to a specified angular momentum order
$L$. HOBCs have three major approximations: (1) nonlinear effects are omitted;
(2) a truncation order $L$ must be chosen; (3) the RWZ formalism is applied in
the BH's rest frame, usually different from the Cauchy frame. Although $\Psi_0$
is an invariant at the linear level, it is still represented on the Cauchy
grid and thus the angular coordinate mismatch at the
boundary between the rest and Cauchy
frames can cause mode mixing. Our CCM implementation relaxes these
approximations, effectively providing an infinite-order $(L=\infty)$ nonlinear
boundary condition.
When linearized around a Schwarzschild background, the CCM algorithm should
reproduce HOBCs.

Finally, the gauge subset determines the remaining four components of $u_{\mu\nu}^{1-}$. It implicitly
sets the boundary conditions for the GH gauge
\cite{Rinne:2006vv, Rinne:2007ui, Rinne:2008vn, Dailey:2024kjg}
\begin{align}
    \Box x^\mu =H^\mu,
\end{align}
where $\Box$ is the d'Alembert operator, and $H^\mu$ are freely specifiable
gauge source functions, here chosen as the  Damped Harmonic gauge
\cite{Szilagyi:2009qz}. While this subset controls the dynamics of the Cauchy
grids, it $\textit{does not}$ affect GWs \cite{Sun:2025yyds} once transformed into the Bondi-Sachs
frame (modulo BMS transformations
\cite{1962RSPSA.269...21B,sachs1962gravitational}), which represents inertial
observers at future null infinity. Therefore, for waveform modeling, CCM is not
required for this subset.

However, this does not mean that the gauge subset is unimportant in numerical
simulations. As noted in \cite{Szilagyi:2015rwa}, poorly chosen gauge boundary
conditions can cause an exponential drift of a binary’s center of mass over a long
timescale. Although this drift is a gauge effect, it can lead to numerical
challenges and increase computational costs. In the following discussions, we
mainly adopt the Sommerfeld boundary condition for the
subset~\cite{Rinne:2007ui}.

To summarize, the primary implementation challenge of CCM is to model backscattered GWs. In strong-gravity regions, it is well known that there is no canonical definition for local physical quantities like gravitational energy. This suggests that CCM could naturally be a cumbersome procedure, requiring the consideration of all dynamical variables. However, the GH formalism simplifies this significantly by providing a framework to isolate the relevant subset of variables. By focusing exclusively on this reduced subset, along with the tricks described in Paper I, CCM becomes computationally manageable.

\section{Numerical simulations}
\label{sec:numerical_simulations}
In this section, we present nine CCM simulations of BBH mergers, with configurations summarized in Table \ref{table:NR_runs}.
Our code setup follows a hybrid approach: we adopt the NR code \texttt{SpECTRE} \cite{deppe_2024_10967177} to handle the characteristic system, while leveraging the more mature code \texttt{SpEC}
\cite{SpECwebsite} for the Cauchy evolution.
Our design links the \texttt{SpECTRE} characteristic module to
\texttt{SpEC}'s executable, enabling \texttt{SpEC} to invoke \texttt{SpECTRE}
functions as needed throughout the simulation. Specifically, we adopt
\texttt{SpEC}'s time stepper, chosen as the fifth-order Dormand-Prince
integrator, to evolve both Cauchy and characteristic variables. The
\texttt{SpECTRE} module is used only for evaluating the right-hand sides of the
characteristic equations and computing quantities necessary for CCM.

Our CCM system can stably evolve BBH mergers without requiring fine-tuning of
constraint damping parameters.  We use common values as in other SXS
simulations, see Eqs.~(53), (54), and Table 2 of
\cite{Lovelace:2024wra}. Simulations are run at three resolutions ( ``Low'',
``Medium'', and ``High''). The Cauchy resolution is set by specifying numerical
error tolerances for the adaptive mesh refinement algorithm
\cite{Szilagyi:2014fna}, while the characteristic resolution is controlled by
the number of grid points in angular and radial directions.

The \texttt{SpEC} initial data solver generates Cauchy initial data through the
Extended Conformal Thin Sandwich formulation \cite{York:1998hy, Pfeiffer:2002iy,
Pfeiffer:2004nc}. Superposed harmonic-Kerr~\cite{Varma:2018sqd} data is adopted
to reduce junk radiation.

While Sec.~II~E of \cite{Moxon:2021gbv} lists various methods to construct
characteristic initial data, it omits details of  \texttt{SpECTRE}'s default algorithm, which we summarize in Appendix \ref{app:characteristic_id}.

\subsection{Head-on collisions}
\label{subsec:runs_head_on}
We first consider head-on collisions of two nonspinning BHs with various mass ratios. As summarized in Table \ref{table:NR_runs}, the initial separation is set to 110$M$, and the initial boost velocity is 0. The outer boundary and the time-like
worldtube
for CCM are placed at a radius of $650M$ from the center of
mass. In this setup, the Cauchy domain is relatively small, making the interior dynamics in causal
contact with the outer boundary within the merger time $(t\sim 1323.7M)$. Consequently, the Cauchy boundary conditions impact the evolution.

\begin{figure}[h]
        \includegraphics[width=\linewidth]{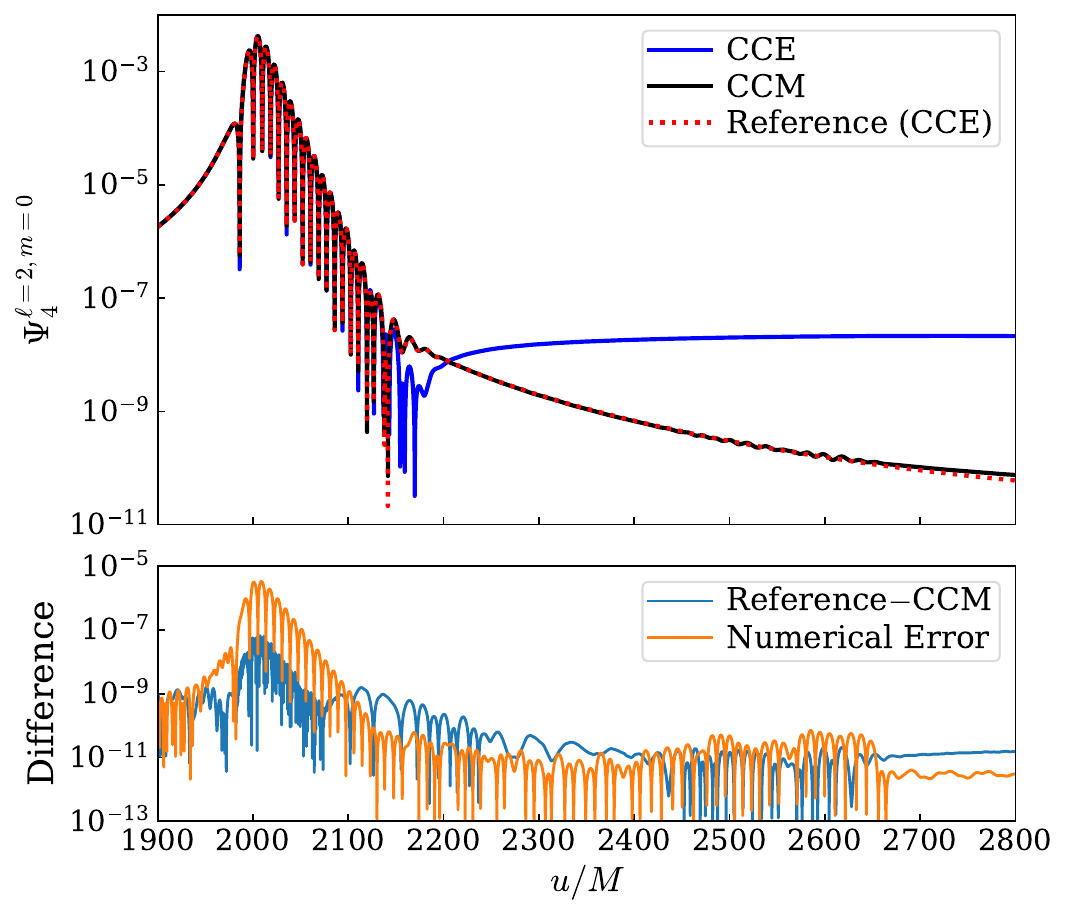}
    \caption{Top panel: $\Psi_4^{\ell=2,m=0}$ emitted from the equal-mass $(q=1)$ head-on BBH collision, simulated using CCE (blue) and CCM (black). The ``High'' resolution is used. They are compared to a reference system (red), whose outer boundary remains causally disconnected from the binary throughout the simulation. Bottom panel: The difference between the reference and CCM results (blue), along with an estimate of the numerical error (orange) by taking the difference between two adjacent CCM resolutions (``Medium'' and ``High'').}
    \label{fig:head_on_psi4_comparison}
\end{figure}

\begin{figure}[!h]
        \includegraphics[width=\linewidth]{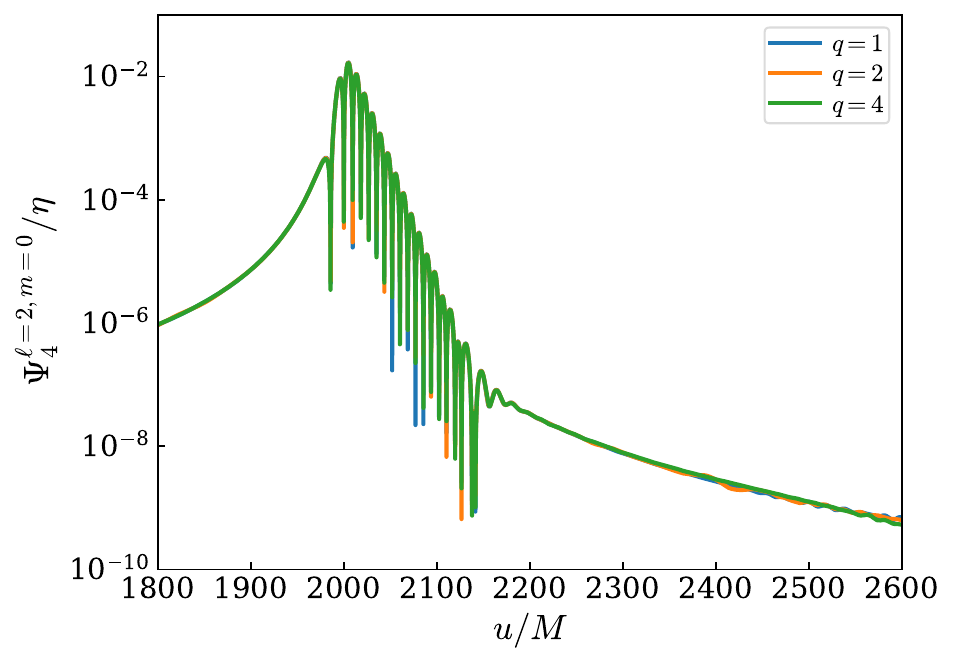}
    \caption{CCM-extracted $\Psi_4^{\ell=2,m=0}$,  normalized by the symmetric mass ratio $\eta$ [Eq.~\eqref{eq:symmetric_mass_ratio}], for the head-on collisions with various mass ratios ($q=1,2,4$; see Table \ref{table:NR_runs}). }
    \label{fig:head_on_mass_ratio}
\end{figure}

The black curve in Fig.~\ref{fig:head_on_psi4_comparison} shows the CCM evolution of
the $(\ell=2,m=0)$ harmonic of $\Psi_4$ at future null infinity for the equal-mass system $(q=1)$, computed at the
highest resolution. It is compared to the extraction without matching (in
blue). We focus on $\Psi_4$ because its functional form is unaffected by
supertranslations, see Eq.~(17e) in \cite{Boyle:2015nqa}, thus avoiding a constant
offset from memory effects\footnote{In other words, $\Psi_4$ is expected to vanish after the system settles, making it more convenient to look for late-time tails.}. Significant differences are evident at late
times. To validate the CCM simulation, we conduct a reference simulation without
CCM, whose outer boundary, positioned at 6000$M$, remains causally disconnected
from the system. This reference result (in red) nearly overlaps the black
curve. The lower panel of Fig.~\ref{fig:head_on_psi4_comparison} shows the
difference between the reference and CCM results (in blue), together with an
estimate of numerical error obtained by taking the difference between two
adjacent CCM resolutions (``Medium'' and ``High'').
The comparable differences confirm that CCM accurately converges to the exact infinite domain problem.

To examine the effect of mass ratio $q$, we normalize $\Psi_{4}^{\ell=2,m=0}$ by the symmetric mass ratio $\eta$:
\begin{align}
    \eta\equiv \frac{m_1 m_2}{(m_1+m_2)^2} \equiv \frac{q}{(1+q)^2}. \label{eq:symmetric_mass_ratio}
\end{align}
As shown in Fig.~\ref{fig:head_on_mass_ratio}, the normalized $\Psi_4$'s align remarkably well across different mass ratios, indicating that the overall time evolution scales cleanly with $\eta$. In fact, this scaling behavior has been previously observed in the perturbative regime (e.g., \cite{DeAmicis:2024eoy}); here, we confirm its validity in the fully nonlinear setup.

\begin{figure}[h]
    \centering
    \includegraphics[width=0.5\textwidth,clip=true]{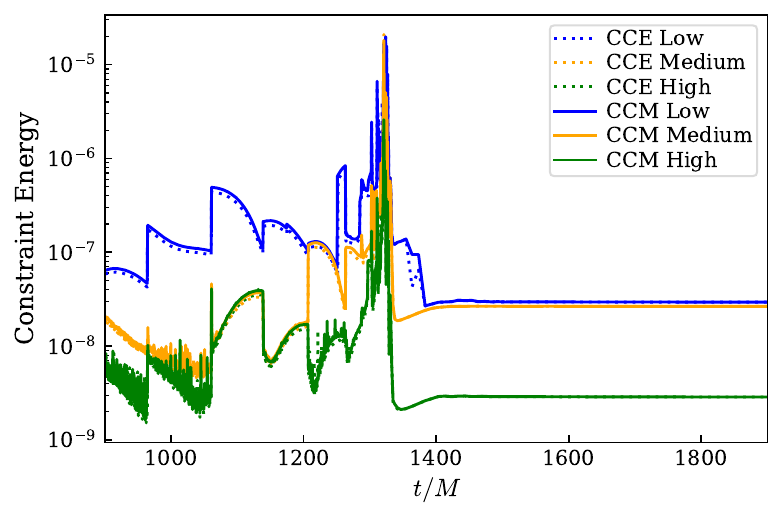}
   \caption{$L^2-$norm of the Cauchy GH constraint energy for the equal-mass head-on collision, simulated with CCM (solid curves) and CCE (dashed curves) at three resolutions. The Cauchy time $t$ is used.}
   \label{fig:cauchy_constraint_energy}
 \end{figure}

 \begin{figure}[h]
    \centering
    \includegraphics[width=0.5\textwidth,clip=true]{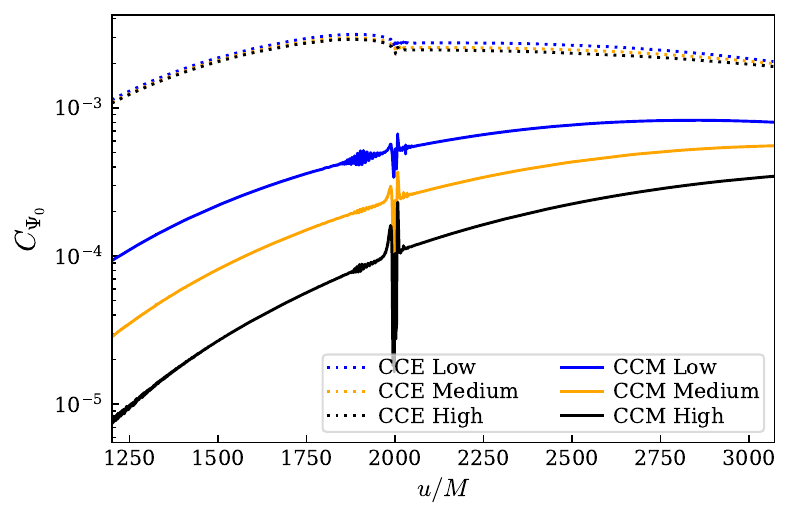}
    \includegraphics[width=0.5\textwidth,clip=true]{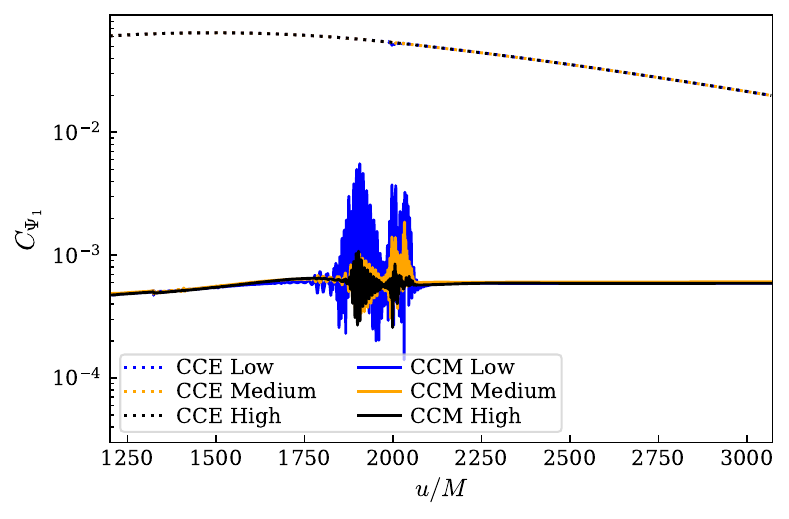}
    \includegraphics[width=0.5\textwidth,clip=true]{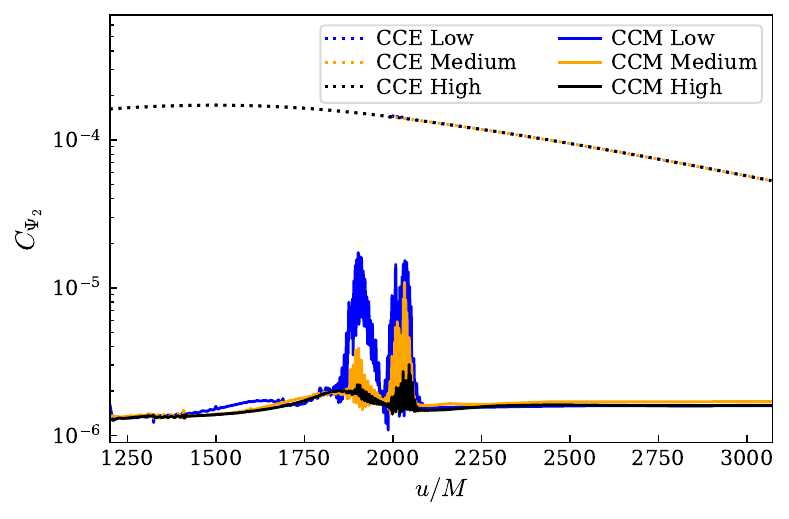}
   \caption{$L^2-$norm of $C_{\Psi_0},C_{\Psi_1}$, and $C_{\Psi_2}$, as defined in Eq.~\eqref{eq:bianchi_identities}, for the equal-mass head-on collision, simulated with CCM (solid curves) and CCE (dashed curves) at three resolutions. The retarded time $u$ is used.}
   \label{fig:bondi_constraints_headon}
 \end{figure}

Besides waveform comparisons, we also assess the accuracy of our CCM simulations by monitoring constraint violations. A CCM system consists of a Cauchy sector and a characteristic sector, each with its own constraints. For the Cauchy sector, constraints are encoded in the GH constraint energy, as defined in Eq.(53) of \cite{Lindblom:2005qh}. Figure~\ref{fig:cauchy_constraint_energy} shows its $L^2-$norm for the equal-mass system at three resolutions (solid curves). The constraints converge with resolution and 
become constant once the system settles down. No
instabilities are observed. A
comparison with a standard Cauchy evolution (without matching but under
otherwise identical conditions) shows that CCM does not significantly alter the
constraint energy (dotted curves), in agreement with Paper I.

As for the characteristic sector, Einstein’s equations and the Bianchi
identities link the Weyl scalars $\Psi_{0,1,2,3,4}$ and the strain $h$ at future null infinity. Specifically, the following constraints should vanish in the exact limit (see, e.g., \cite{Ashtekar:2019viz, Iozzo:2020jcu}):
\begin{subequations}
    \label{eq:bianchi_identities}
    \begin{align}
    &C_{\Psi_2}\equiv\dot{\Psi}_2+\frac{1}{2}\eth\Psi_3-\frac{1}{4}\bar{h}\Psi_4  ,  \\
    &C_{\Psi_1}\equiv\dot{\Psi}_1+\frac{1}{2}\eth\Psi_2-\frac{1}{2}\bar{h}\Psi_3  ,\\
    &C_{\Psi_0}\equiv\dot{\Psi}_0+\frac{1}{2}\eth\Psi_1-\frac{3}{4}\bar{h}\Psi_2,
\end{align}
\end{subequations}
where the dot represents the retarded-time derivative, $\eth$ denotes the angular derivative, and the bar stands for complex conjugation. Since our characteristic evolution computes those quantities independently, these relations yield nontrivial constraints that can be used to assess the accuracy of our simulations.
Figure~\ref{fig:bondi_constraints_headon} shows the $L^2-$norm of $C_{\Psi_0,\Psi_1,\Psi_2}$ for the equal-mass system. We can see they converge with resolution.
In contrast, waveforms extracted using CCE from standard Cauchy evolutions
without matching (dotted lines) show much poorer convergence.

Finally, we evaluate the numerical efficiency of our CCM code. Since the characteristic system is more efficient in handling the wave zone, CCM is expected to improve computational efficiency by reducing the size of the Cauchy domain \cite{winicour2012characteristic}. Table~\ref{table:NR_runs} summarizes the CPU hours used by the CCM simulations and their corresponding reference systems. Even without full optimization, all three CCM simulations are already $1.8-3$ times more efficient.

\subsection{Quasi-head-on collisions}
\label{subsec:quasi_head_on}
Our second class of simulations includes four quasi-head-on collisions of equal-mass nonspinning BHs. As summarized in Table \ref{table:NR_runs}, we systematically increase the initial angular velocity $\Omega_{\rm orb}$ (the transverse velocity relative to the line connecting the two BHs), while fixing the initial radial velocity at 0. In the special scenario where $\Omega_{\rm orb}=0$, the system reduces to the purely head-on scenario studied in Sec.~\ref{subsec:runs_head_on}.

\begin{figure}[!h]
        \includegraphics[width=\linewidth]{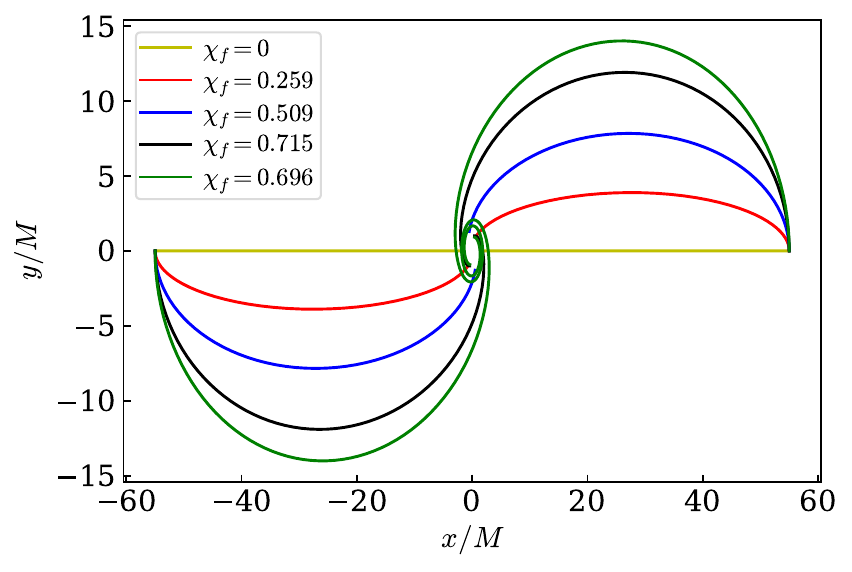}
    \caption{Trajectories of the BHs in the $x-y$ plane for the (quasi-)head-on systems listed in Table \ref{table:NR_runs}. The initial angular velocity is along the $z-$axis. Each system is labeled by the dimensionless spin $(\chi_f)$ of the remnant Kerr BH. }
    \label{fig:quasi_headon_traj}
\end{figure}

\begin{figure*}[!htb]
    \subfloat[$\chi_f=0.259$]{\includegraphics[width=0.49\linewidth]{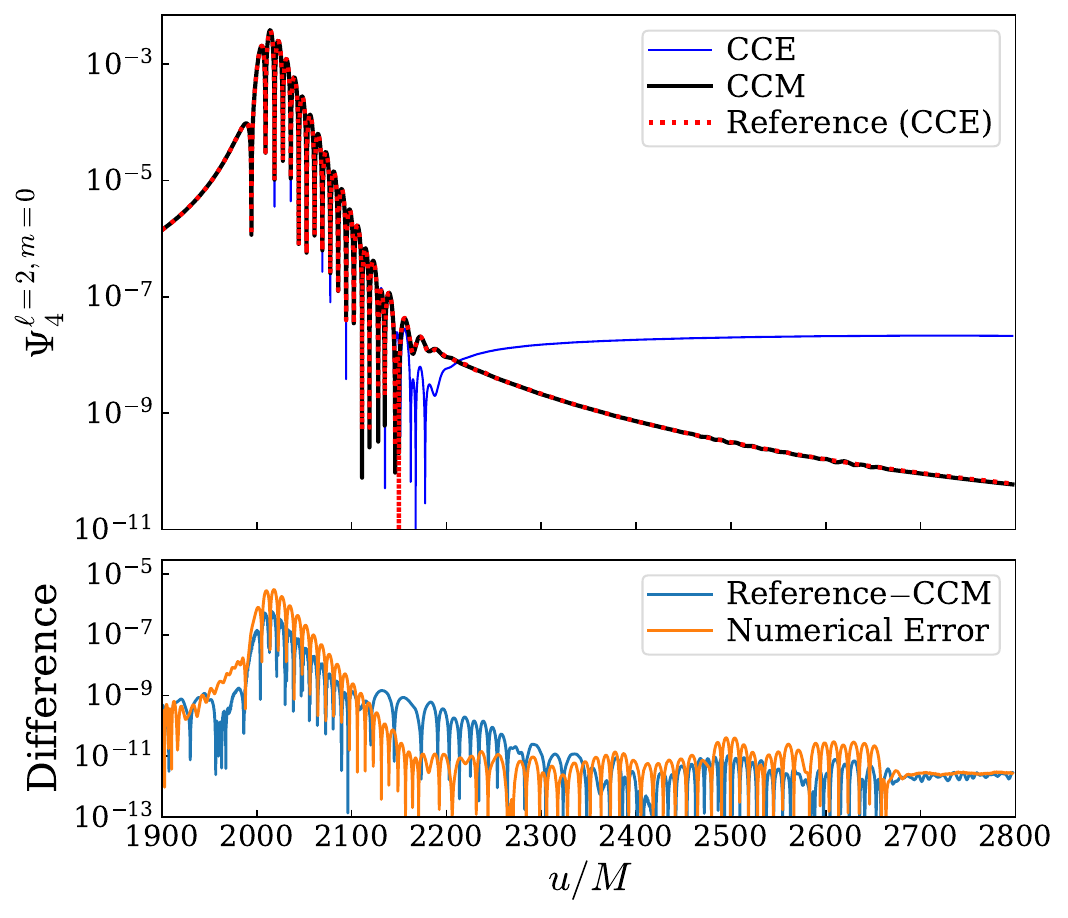}}
    \subfloat[$\chi_f=0.509$]{\includegraphics[width=0.49\linewidth]{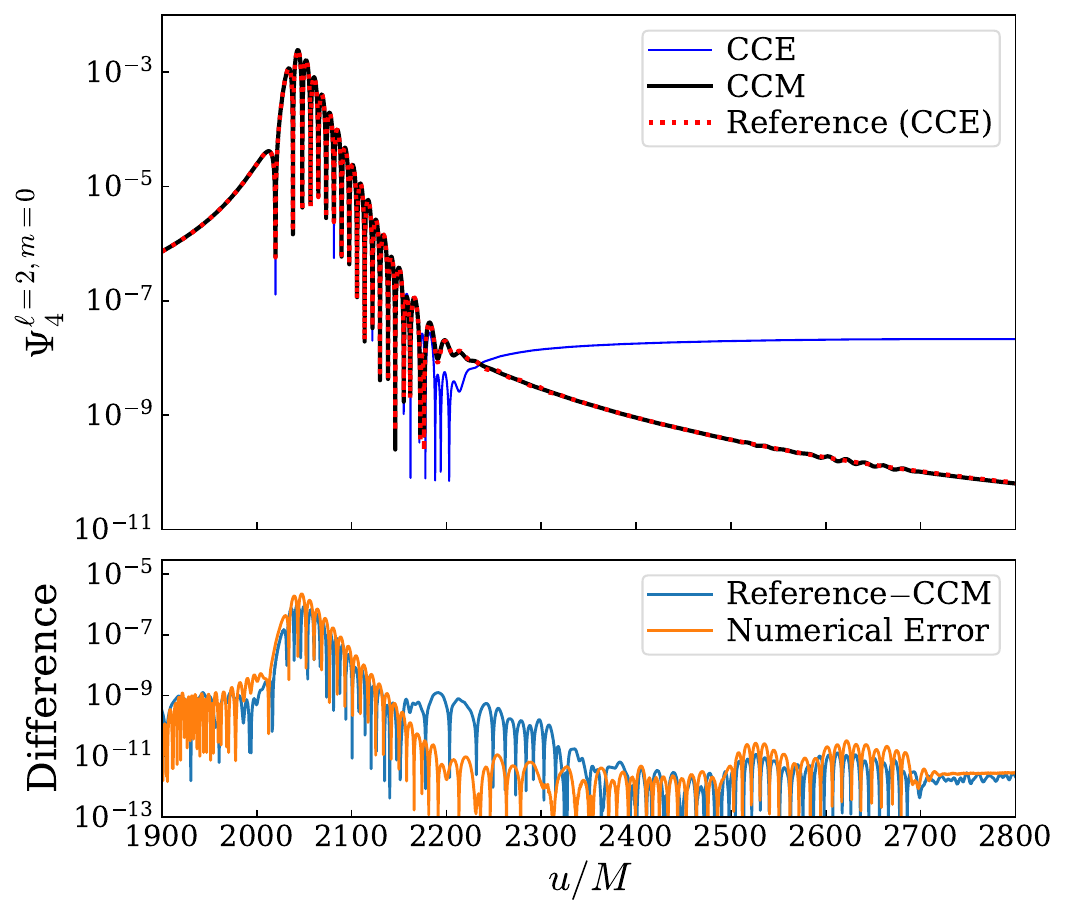}} \\
    \subfloat[$\chi_f=0.715$ \label{fig:point3_psi4}]{\includegraphics[width=0.49\linewidth]{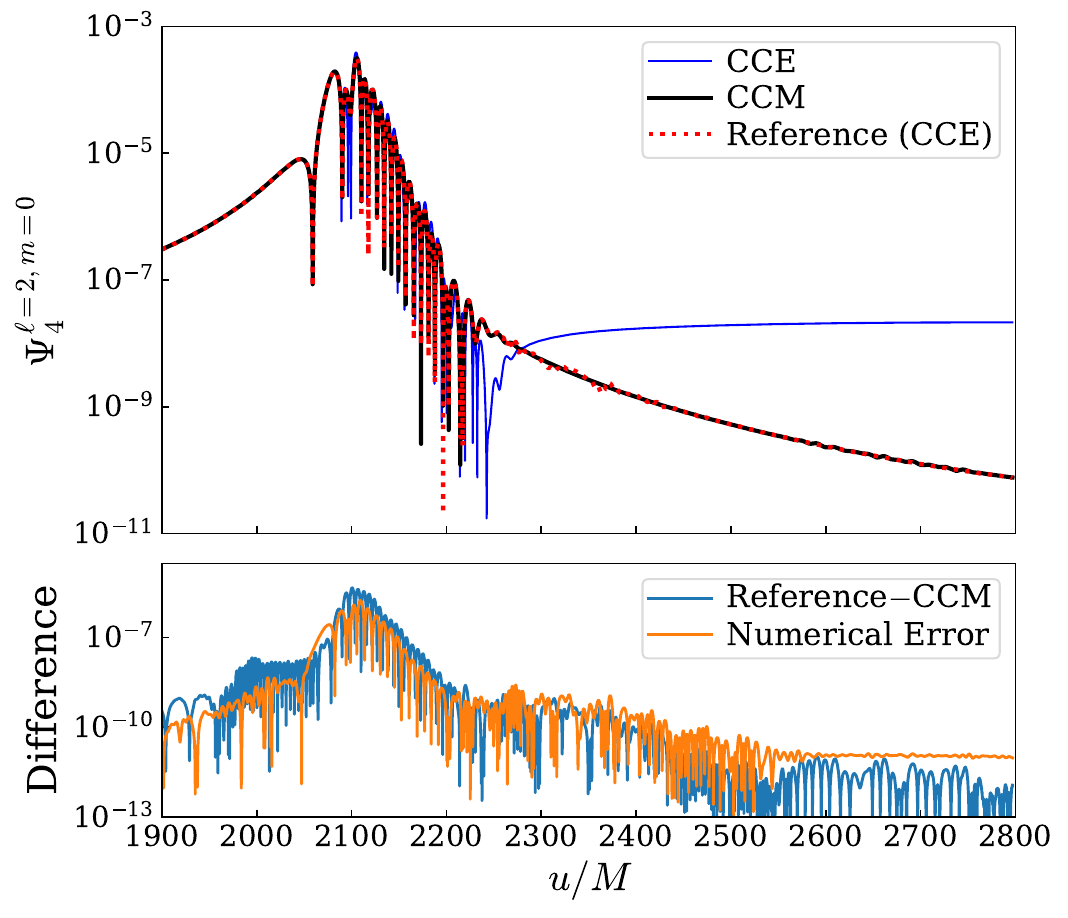}}
    \subfloat[$\chi_f=0.696$\label{fig:point35_psi4}]{\includegraphics[width=0.49\linewidth]{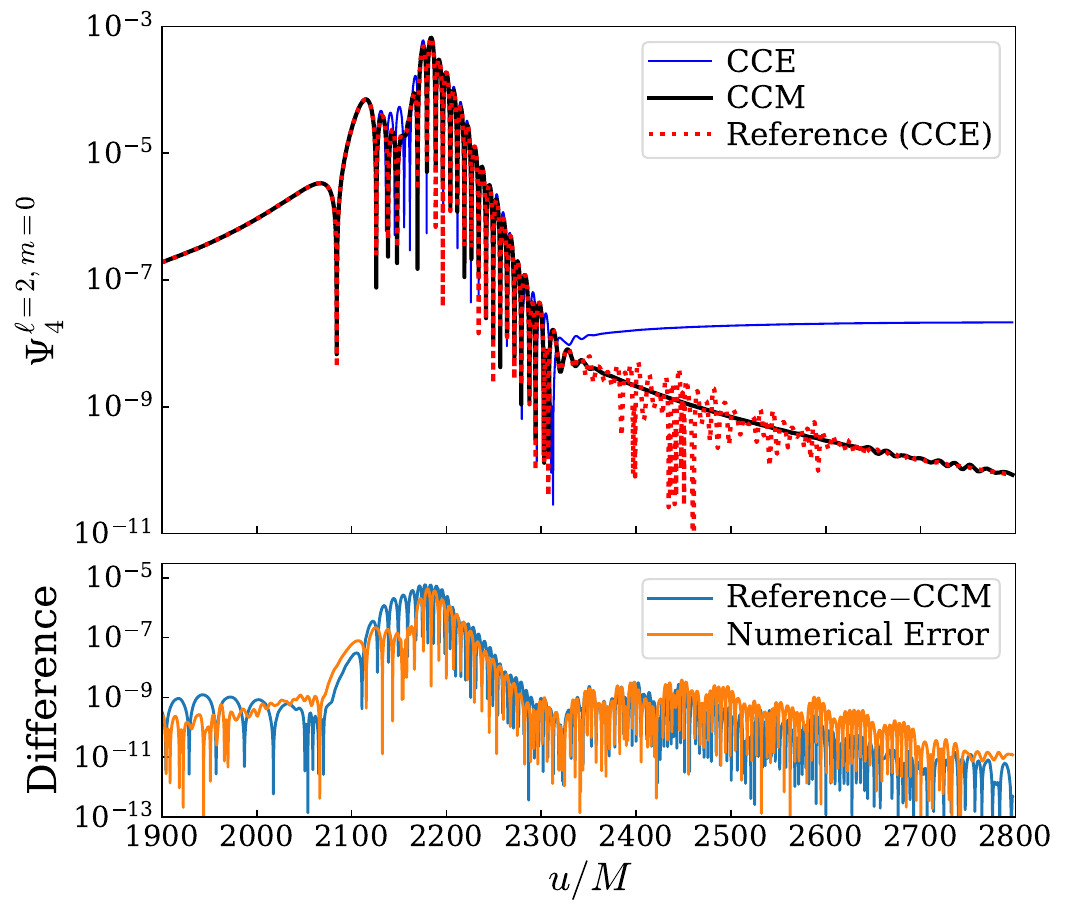}}
\caption{Similar to Fig.~\ref{fig:head_on_psi4_comparison}, comparison of $\Psi_4^{\ell=2,m=0}$ extracted from CCE, CCM, and reference systems for the quasi-head-on collisions listed in Table \ref{table:NR_runs}. The initial angular velocity $\Omega_{\rm orb}$ increases from (a) to (d), but the dimensionless spin of the remnant Kerr BHs $(\chi_f)$ does not increase monotonically --- more angular momentum is radiated away via GWs in Fig.~\ref{fig:point35_psi4} than in \ref{fig:point3_psi4}. Additionally, high$-\Omega_{\rm orb}$ systems (Figs.~\ref{fig:point3_psi4} and \ref{fig:point35_psi4}) exhibit pre-ringdown emission.}
\label{fig:quasi_head_on_psi4_comparison}
\end{figure*}

Figure \ref{fig:quasi_headon_traj} shows the trajectories of the BHs in the $x-y$ plane (the initial angular velocity is along the $z-$axis). As $\Omega_{\rm orb}$ increases, the BHs undergo a partial orbital motion before merging. Here we ensure that $\Omega_{\rm orb}$ remains small to prevent an eccentric orbit.
The remnants of these collisions are Kerr BHs, whose dimensionless spins ($\chi_f$) are listed in Table \ref{table:NR_runs}. Notably, $\chi_f$ does not increase monotonically with $\Omega_{\rm orb}$, as a larger fraction of angular momentum may be radiated away in GWs during the inspiral phase. Below, for convenience, we use $\chi_f$ to label these quasi-head-on systems.

Figure \ref{fig:quasi_head_on_psi4_comparison} compares $\Psi_{4}^{\ell=2,m=0}$ extracted from CCM (in black), CCE (in blue), and reference systems (in red). Once again, we observe noticeable improvements in the CCM waveforms compared to CCE. 
In particular, the cases shown in Figs.~\ref{fig:point3_psi4} and \ref{fig:point35_psi4} have relatively large initial angular velocities, leading to pre-ringdown GW emission. 
Their reference results display numerical wiggles in the late-time tail, whereas the CCM waveforms (using the same numerical resolution) remain accurate. Notably, as summarized in Table \ref{table:NR_runs}, the CCM simulations require $2-3$ times fewer CPU hours than the reference runs. These comparisons demonstrate that our CCM algorithm improves both waveform accuracy and computational efficiency.

\begin{figure*}[!htb]
        \includegraphics[width=0.49\linewidth]{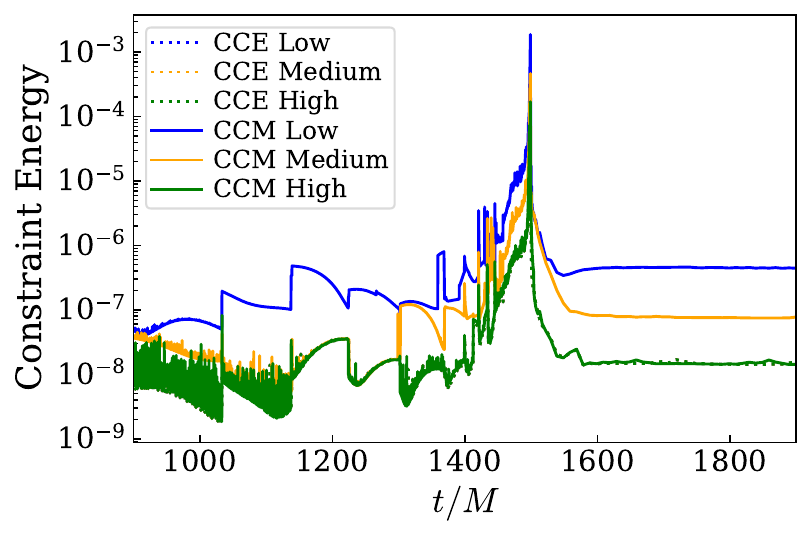}
        \includegraphics[width=0.49 \linewidth]{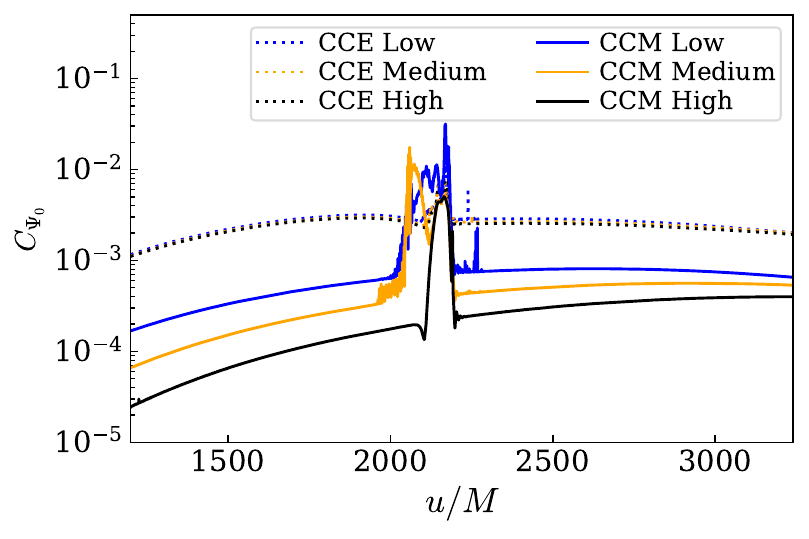} \\
        \includegraphics[width=0.49 \linewidth]{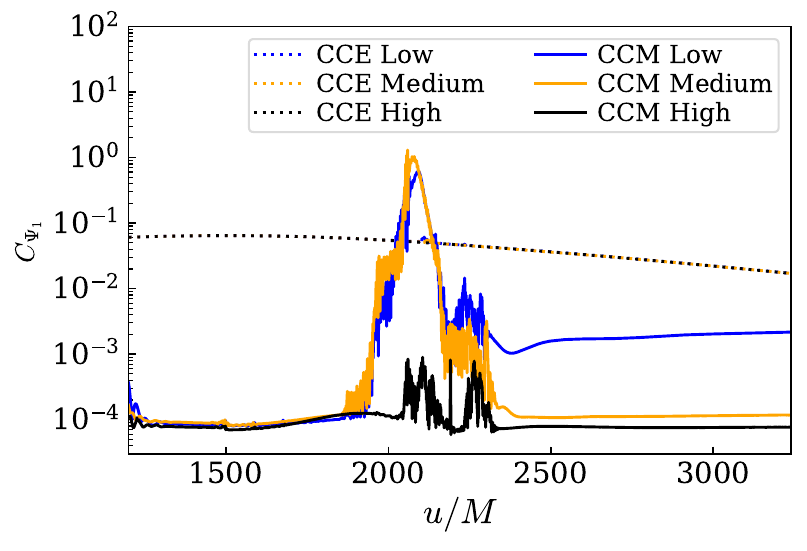}
        \includegraphics[width=0.49 \linewidth]{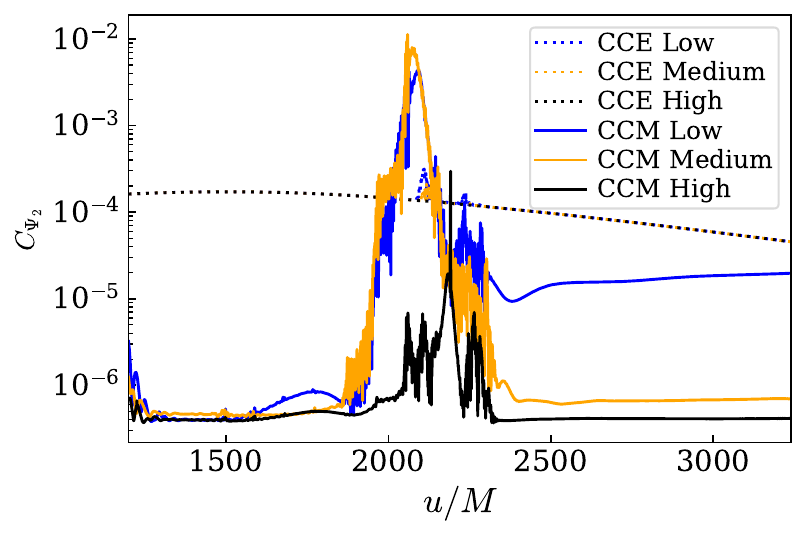}
    \caption{Similar to Figs.~\ref{fig:cauchy_constraint_energy} and \ref{fig:bondi_constraints_headon}, $L^2-$norm of the Cauchy GH constraint energy (top left) and $C_{\Psi_0},C_{\Psi_1}$, $C_{\Psi_2}$ for the system in Fig.~\ref{fig:point35_psi4}, using CCM (solid curves) and CCE (dotted curves).  }
    \label{fig:point35_constraints}
\end{figure*}

Finally, we monitor the constraint violations throughout the simulations. As a representative case, we select the system in Fig.~\ref{fig:point35_psi4}, which has the largest initial angular velocity and the most complex pre-merger dynamics. Figure \ref{fig:point35_constraints} provides the $L^2$-norm of its GH constraint energy and $C_{\Psi_0,\Psi_1,\Psi_2}$. The results are consistent with our findings in Sec.~\ref{subsec:runs_head_on} and Paper I: (1) The CCM constraints converge with resolution. (2) CCM improves the characteristic constraints while keeping the GH constraint energy comparable to the systems without matching.

\subsection{Eccentric binary}
\label{subsec:eccentric}
Our CCM simulations in Secs.~\ref{subsec:runs_head_on} and \ref{subsec:quasi_head_on} are relatively short and have simple pre-merger dynamics. To probe the long-term behavior of the algorithm, we now consider an eccentric binary system with randomly chosen initial angular and radial velocities, as listed in Table~\ref{table:NR_runs}. The initial separation is set to $90M$. 

\begin{figure}[htb]
        \includegraphics[width=\linewidth]{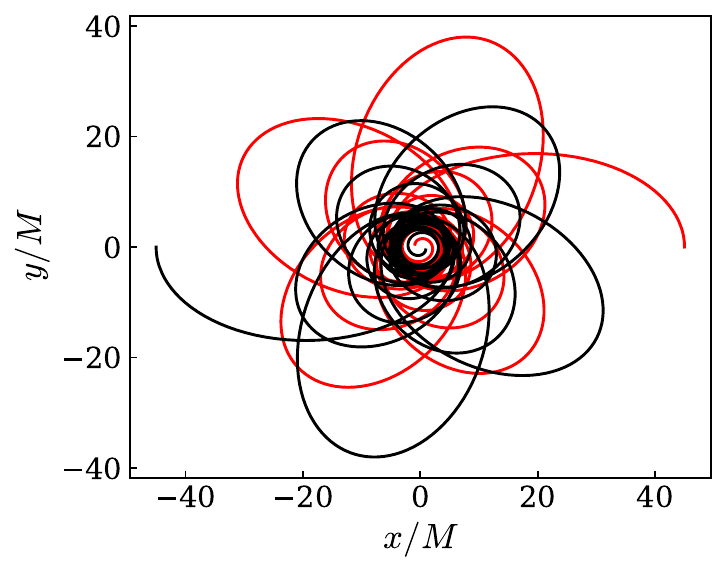}
    \caption{Trajectories of the BHs in the $x-y$ plane for the eccentric system listed in Table \ref{table:NR_runs}. The initial angular velocity is along the $z-$axis. }
    \label{fig:ecc_traj}
\end{figure}

Figure \ref{fig:ecc_traj} shows the trajectories of the two component BHs. The orbit begins with a large eccentricity $(>0.8)$ and exhibits strong perihelion precession. Over the $22.4$ orbital cycles ($\sim44.8$ GW cycles) before the merger, the orbit gradually circularizes due to GW emission. The merger time is $14781M$, roughly $10$ times longer than those studied in Secs.~\ref{subsec:runs_head_on} and \ref{subsec:quasi_head_on}. Given the long duration of the simulation, evolving the corresponding reference system becomes computationally impractical.

\begin{figure}[!htb]
        \includegraphics[width=\linewidth]{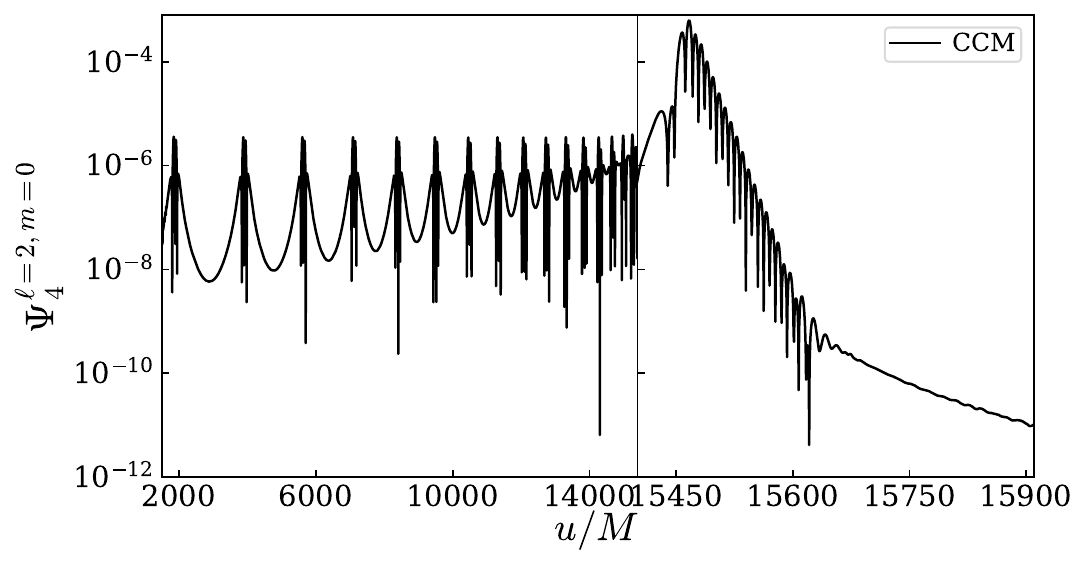} \\
        \includegraphics[width=\linewidth]{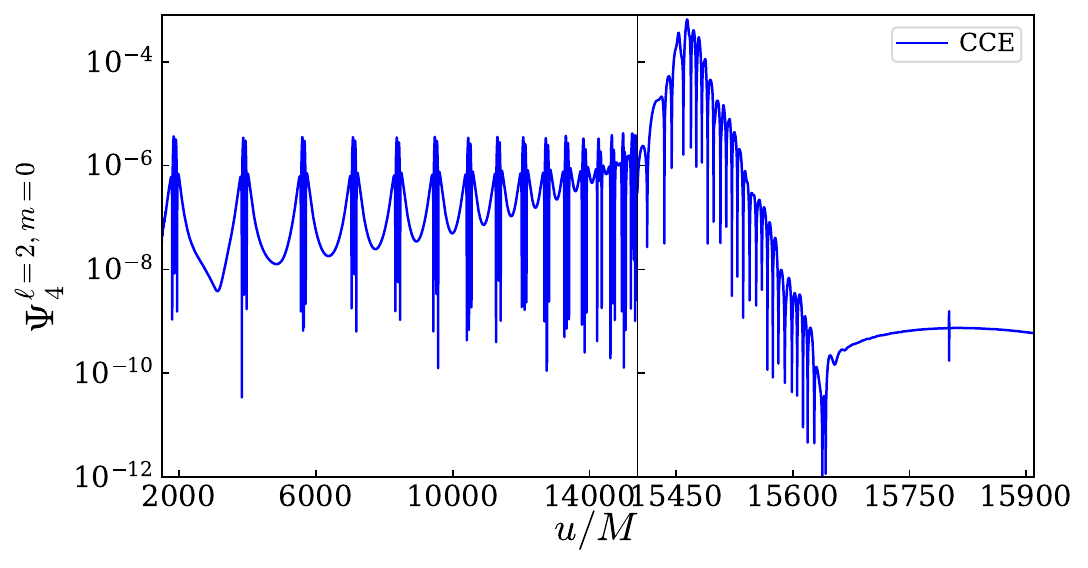}
    \caption{Inspiral-merger-ringdown $\Psi_4^{\ell=2,m=0}$ extracted using CCM (top panel) and CCE (bottom panel) from the eccentric system listed in Table \ref{table:NR_runs}.  }
    \label{fig:eccentric_psi4}
\end{figure}

As in previous cases, no instabilities are observed in the CCM evolution. Figure \ref{fig:eccentric_psi4} shows the full inspiral-merger-ringdown waveform $\Psi_4^{\ell=2,m=0}$ computed using CCM (in black) and CCE (in blue). During the inspiral stage, sharp bursts of GW radiation are emitted near each perihelion passage. Following the ringdown, a late-time tail is clearly visible, but this component is obscured by an error floor in the absence of CCM.

\subsection{Quasi-circular binary}
\begin{figure}[!h]
        \includegraphics[width=\linewidth]{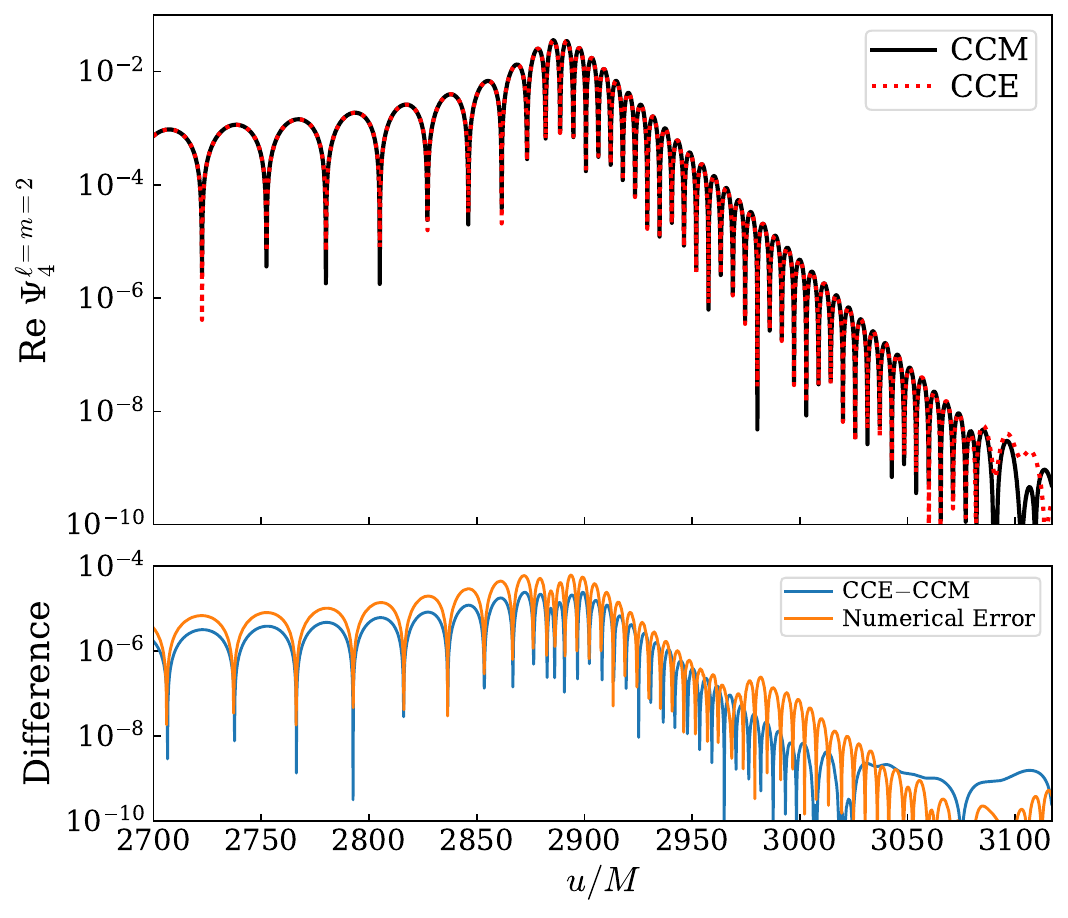}
    \caption{Top panel:  $\Psi_4^{\ell=m=2}$ from a quasi-circular collision, simulated using CCM (black) and CCE (red). Bottom panel: The difference between the two results (blue), compared to the numerical error (orange).}
    \label{fig:imr_psi4}
\end{figure}

Our final case is a merger of two equal-mass, nonspinning BHs on a quasi-circular
orbit. The properties of this system are summarized in Table \ref{table:NR_runs}. The orbital eccentricity is iteratively reduced to below $\sim
4\times10^{-4}$~\cite{Pfeiffer:2007yz, Buonanno:2010yk, Mroue:2012kv}. Figure~\ref{fig:imr_psi4} compares $\Psi_4^{\ell=m=2}$ 
extracted with and without CCM. In this case, the improvement from CCM is negligible, and there is no hint of late-time tails.

%==========================================================================
\section{Tail analysis with rational filters}
\label{sec:tail_analysis}
CCM has opened a systematic path for studying late-time tails in BBH systems. Building on the results in Sec.~\ref{sec:numerical_simulations}, we now present a phenomenological analysis of the tail behavior in the (quasi-)head-on configurations listed in Table \ref{table:NR_runs}.

One challenge in tail extraction is the impact of QNMs, which can mask the tail signal, especially at early times. A common workaround is to delay the analysis window until the QNMs have sufficiently decayed. However, this approach often results in unnecessary signal loss, particularly in NR waveforms. Moreover, the tail component may already be present beneath the QNM ringing at earlier times. The workaround risks overlooking the early-time contribution.

To overcome this, we adopt QNM rational filters \cite{Ma:2022wpv,Ma:2023cwe,Ma:2023vvr} and the \texttt{PYTHON} package \texttt{qnm\_filter} \cite{FilterGithub}. The filters enable mode removal given only their complex frequencies, without requiring knowledge of the corresponding mode amplitudes. This feature is particularly important, as accurate amplitude fitting demands a detailed model of the early-time tail --- precisely the component we aim to isolate. The filters effectively break this circular dependency and provide a cleaner separation of tail and QNM content.

We first apply the Fourier-analysis method from Sec.~III~B of \cite{May:2024rrg} to identify QNMs in $\Psi_{4}^{\ell=2,m=0}$. The extracted modes are listed in Table \ref{table:tail_results}. In particular, we find a few quadratic QNMs.
We follow the labeling conventions of \cite{Khera:2024yrk}. A linear QNM is labeled by four indices $(\ell,m,n,p)$, where $(\ell,m)$ denote the angular harmonic indices; $n$ is the overtone number; and $p={\rm sgn}({\rm Re} \,\omega)$ indicates the sign of the real part of the mode frequency. Two modes with the same $(\ell,m,n)$ but opposite $p$ are the mirror of each other. A quadratic mode is labeled by  $(\ell_1,m_1,n_1,p_1;\ell_2,m_2,n_2,p_2)$, with its complex frequency given by the sum of the constituent linear mode frequencies.

\begin{figure}[!h]
        \includegraphics[width=\linewidth]{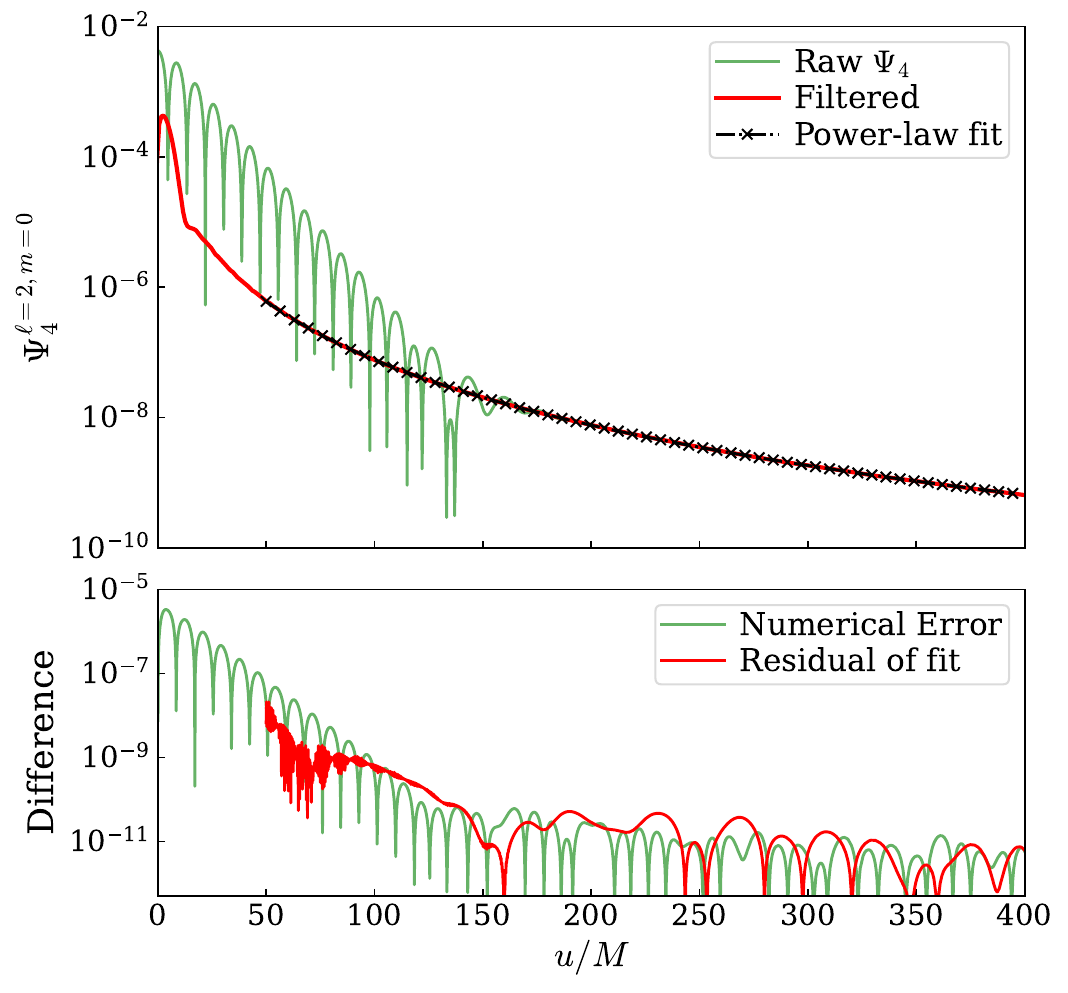}
    \caption{Top panel: The raw (green) and filtered (red) $\Psi_4^{\ell=2,m=0}$ simulated using CCM for the head-on collision of two equal-mass, nonspinning BHs. The power-law fit in Eq.~\eqref{eq:power_law_fit} and Table \ref{table:tail_results} is shown in black (marked with crosses). Bottom panel: The residual of the power-law fit (red), compared to the numerical error (green) by taking the difference between two adjacent CCM resolutions.}
    \label{fig:tail_analysis_with_filter}
\end{figure}

\begin{figure*}[!htb]
        \includegraphics[width=0.49\linewidth]{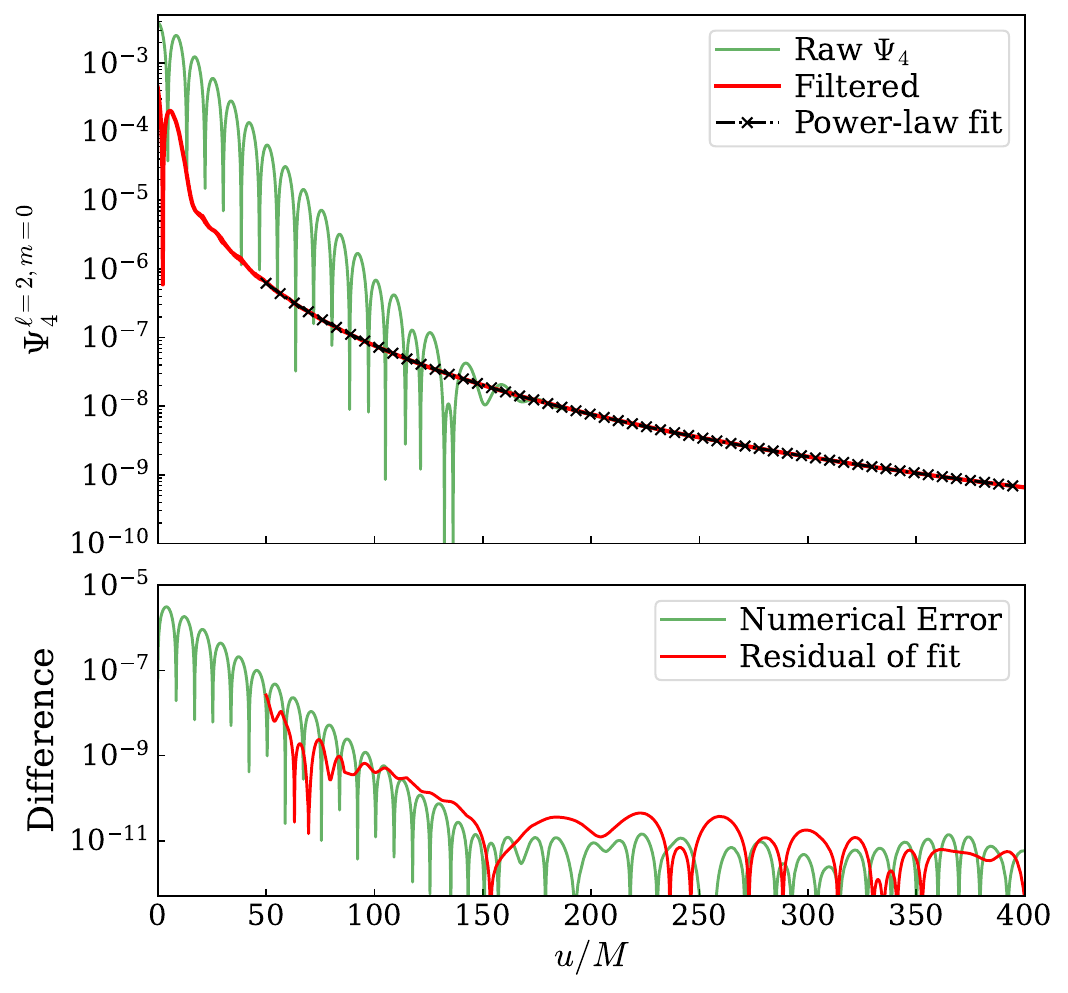}
        \includegraphics[width=0.49\linewidth]{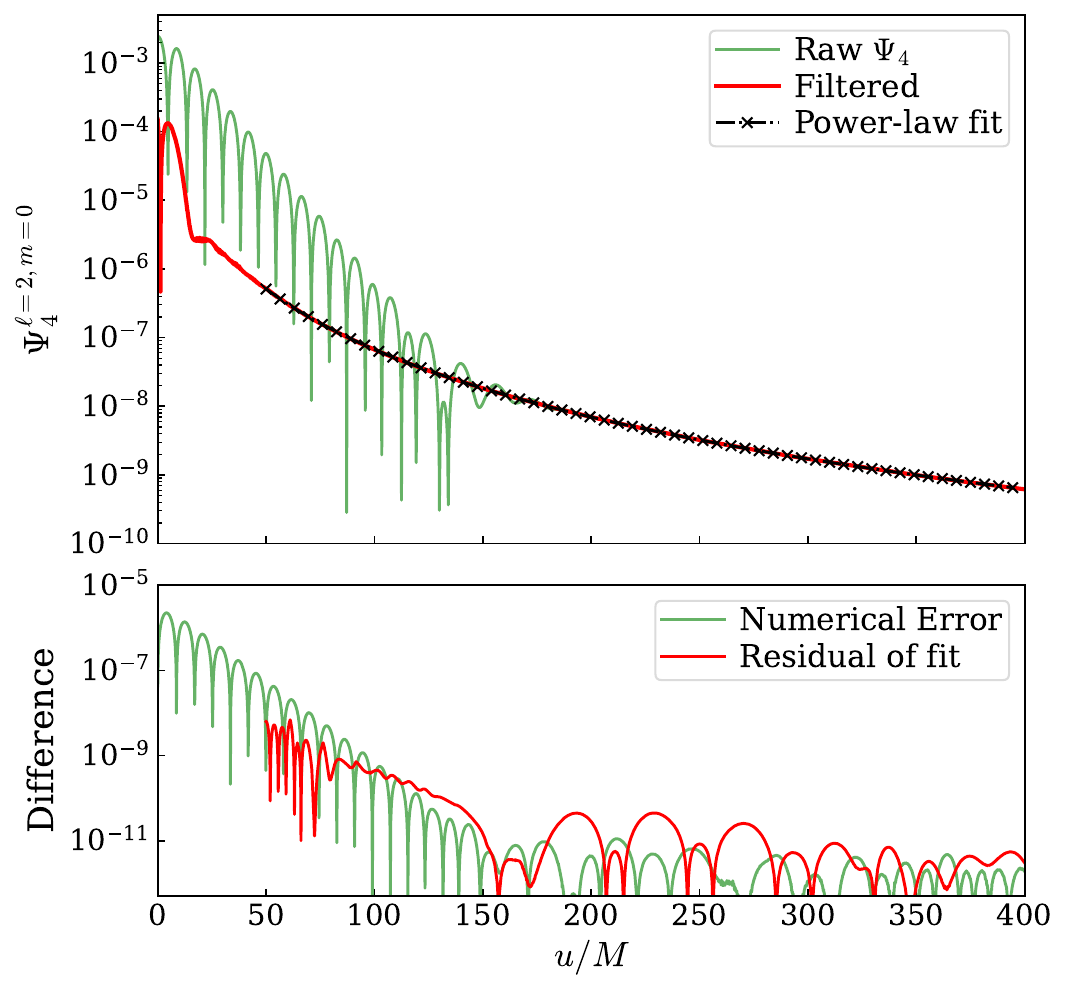} \\
        \includegraphics[width=0.49\linewidth]{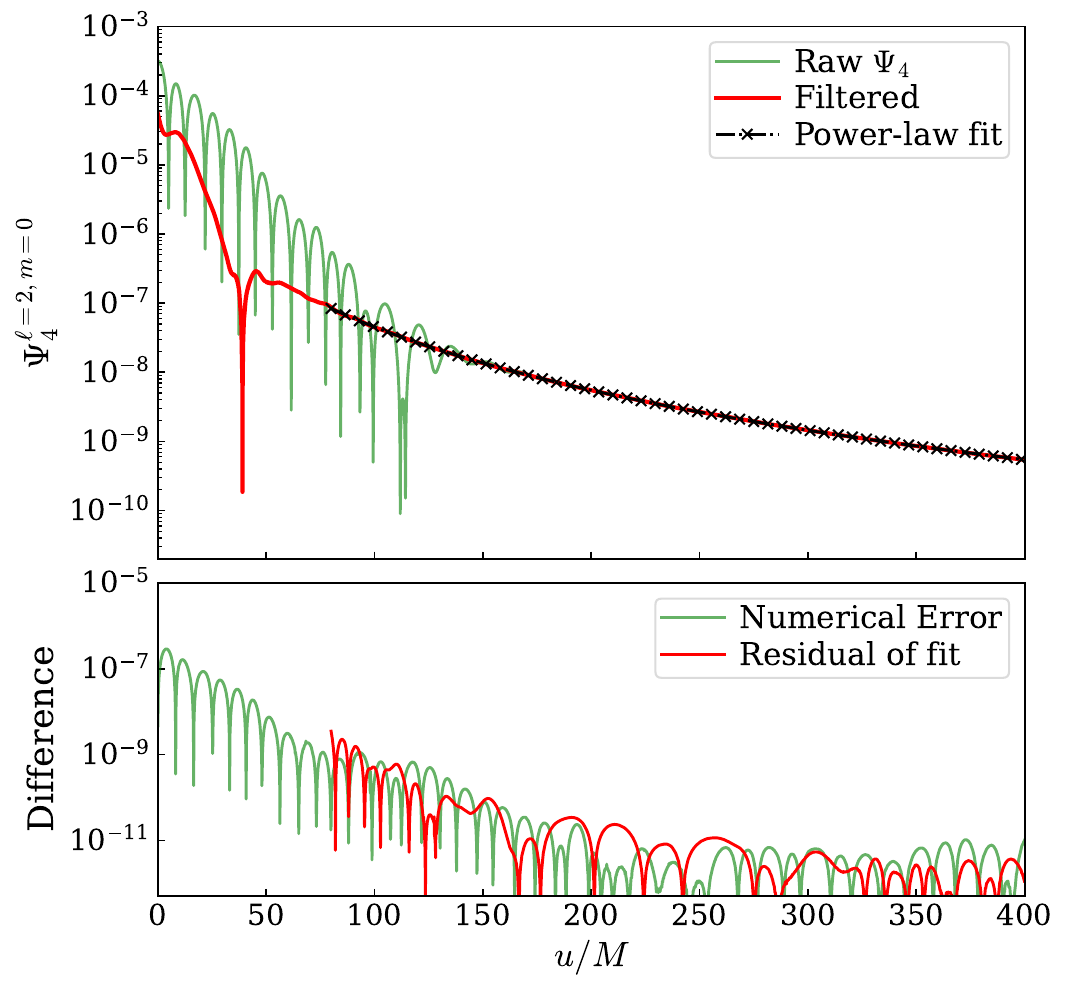} 
        \includegraphics[width=0.49\linewidth]{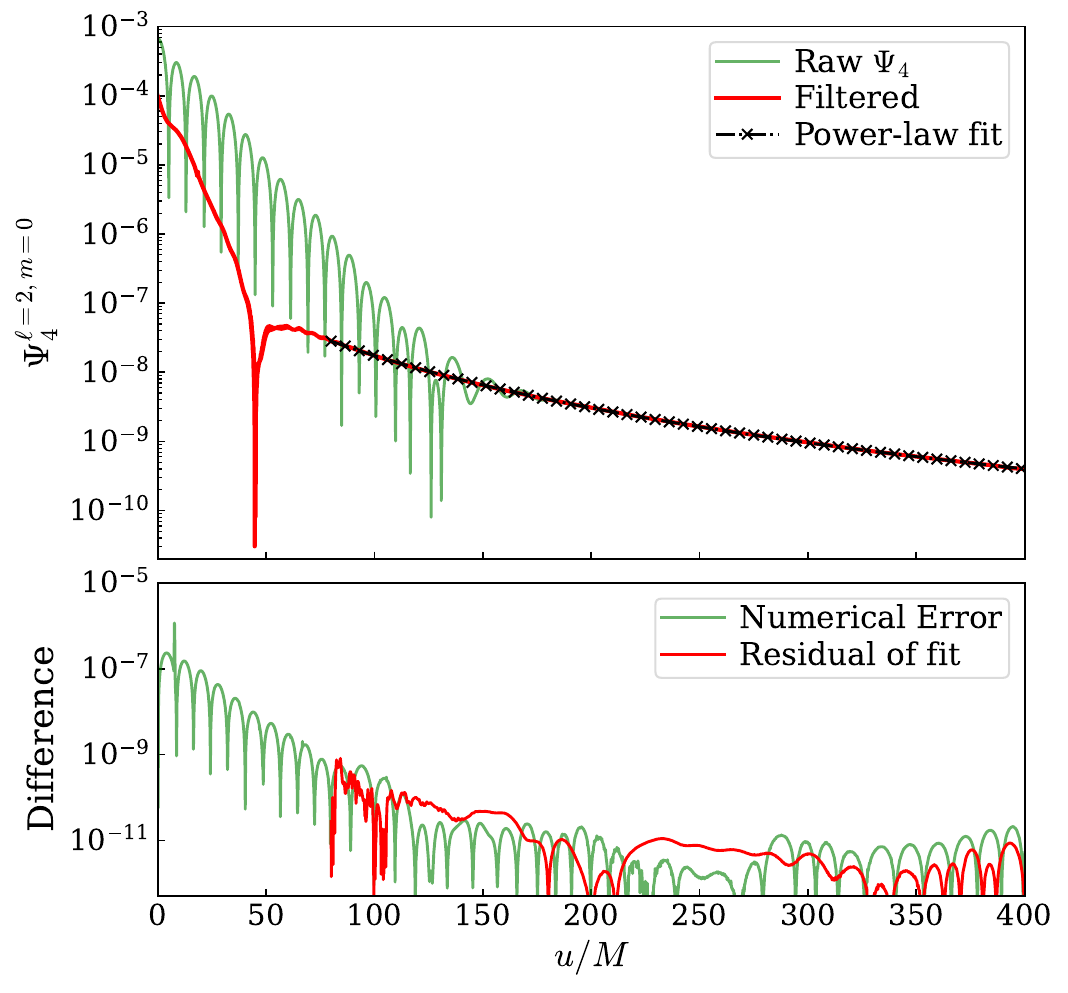} 
    \caption{Similar to Fig.~\ref{fig:tail_analysis_with_filter}, tail analysis for the quasi-head-on binary systems listed in Table \ref{table:NR_runs}. }
    \label{fig:tail_analysis_with_filter_2}
\end{figure*}

After filtering out the QNMs from the ringdown regime, the resulting filtered waveforms are shown as red curves in the upper panels of Figs.~\ref{fig:tail_analysis_with_filter} and \ref{fig:tail_analysis_with_filter_2}. The time axis has been adjusted such that the peak of the original $\Psi_4^{\ell=2,m=0}$ occurs at $u=0$.
As noted in Sec.~II~B and Fig.~3 of \cite{Ma:2022wpv}, these filters introduce a backward time shift in power-law tails, with an analytical expression provided in Eq.~(16) therein.
We have verified that this formula accurately describes our case. Therefore, we apply this formula to adjust for the time shift in the following analysis.

\begin{table*}
    \centering
    \caption{Tail analysis for (quasi-)head-on binaries listed in Table \ref{table:NR_runs}. Each case is labeled by the dimensionless spin $\chi_f$ of the remnant Kerr BH. The Weyl scalar $\Psi_4$ is filtered using rational filters constructed from the corresponding QNM contents, following the labeling conventions of Ref.~\cite{Khera:2024yrk}. The filtered $\Psi_4$ is fit to a single power law model in Eq.~\eqref{eq:power_law_fit}, within a specified time window.}
    \begin{tabular}{c c c c c c c} \hline\hline
     Systems & $\chi_f$ & Fitting Window & $A$ & $u_0$ & $p$  & QNM contents \\ \hline\hline
     \multirow{2}{*}{Head-on} & \multirow{2}{*}{$\sim 0$} & \multirow{2}{*}{$[50M,400M]$} & \multirow{2}{*}{$22.7\eta$} & \multirow{2}{*}{$19.1$} & \multirow{2}{*}{$-3.79$} & $(2,0,0,\pm), (2,0,1,\pm)$ \\ 
     & & & & & &  $(2,0,0,\pm;2,0,0,\pm)$ \\ \hline
     \multirow{8}{*}{\shortstack{Quasi \\Head-on }} & \multirow{2}{*}{$0.259$} & \multirow{2}{*}{$[50M,400M]$} & \multirow{2}{*}{$4.53$} & \multirow{2}{*}{$18.5$} & \multirow{2}{*}{$-3.75$} & $(2,0,0,\pm), (2,0,1,\pm),(2,2,0,\pm),(3,2,0,\pm)$ \\ 
     & & & & & &  $(2,0,0,\pm;2,0,0,\pm),(2,2,0,\pm;2,2,0,\pm)$ \\ \cline{2-7}
     & \multirow{2}{*}{$0.509$} & \multirow{2}{*}{$[50M,400M]$} & \multirow{2}{*}{$3.74$} & \multirow{2}{*}{$19.4$} & \multirow{2}{*}{$-3.73$} & $(2,0,0,\pm), (2,0,1,\pm),(2,2,0,\pm),(3,2,0,\pm)$\\ 
     & & & & & &  $(3,0,0,\pm),(2,0,0,\pm;2,0,0,\pm)$ \\ \cline{2-7}
     & \multirow{2}{*}{$0.715$} & \multirow{2}{*}{$[80M,400M]$} & \multirow{2}{*}{$3.28$} & \multirow{2}{*}{$31.0$} & \multirow{2}{*}{$-3.71$}  & $(2,0,0,\pm), (2,0,1,\pm),(2,2,0,\pm),(3,2,0,\pm)$ \\
    & & & & & &  $(3,0,0,\pm),(2,0,0,\pm;2,0,0,\pm)$ \\ \cline{2-7}
     & \multirow{2}{*}{$0.696$} & \multirow{2}{*}{$[80M,400M]$} & \multirow{2}{*}{0.86} & \multirow{2}{*}{$55.6$} & \multirow{2}{*}{$-3.51$} & $(2,0,0,\pm), (2,0,1,\pm),(2,2,0,\pm),(3,2,0,\pm)$ \\
    & & & & & &  $(3,0,0,\pm),(2,0,0,\pm;2,0,0,\pm)$  \\ \hline\hline
     \end{tabular}
     \label{table:tail_results}
\end{table*}

The filters extend the duration of the nonoscillatory regime, while ensuring that the late-time portion remains accurately aligned with the raw $\Psi_4^{\ell=2,m=0}$ (green curves in the upper panels of Figs.~\ref{fig:tail_analysis_with_filter} and \ref{fig:tail_analysis_with_filter_2}).
Using the fitting windows listed in Table \ref{table:tail_results}, we find that each filtered waveform can be well described by a single power law:
\begin{align}
    A(u+u_0)^p. \label{eq:power_law_fit}
\end{align}
To improve fitting performance, we borrow the spirit of
variable projection \cite{o2013variable,Giesler:2024hcr} by separating linear parameters from
nonlinear ones. This step reduces a multidimensional fitting problem to a 1D
problem. We have checked that the fits are highly stable against the initial
guess. For more details, see the discussion around Eq.~(73) in
\cite{Ma:2024bed}. The residuals of the fits are provided in the lower panels of Figs.~\ref{fig:tail_analysis_with_filter} and \ref{fig:tail_analysis_with_filter_2}, where they are compared to an
estimate of numerical error (the difference between two
adjacent CCM resolutions). The residuals and error estimates are comparable, confirming that the single power-law model reasonably describes the filtered waveforms in the selected time windows.

The extracted fitting parameters $A$, $u_0$, and $p$ are summarized in Table \ref{table:tail_results}. For the head-on cases, we have factored out the symmetric mass ratio $\eta$ [Eq.~\eqref{eq:symmetric_mass_ratio}], motivated by the scaling $\Psi_{4}\propto\eta$ observed in Fig.~\ref{fig:head_on_mass_ratio}. The results show that the tail amplitude $A$ systematically decreases as the initial orbital angular velocity ($\Omega_{\rm orb}$ in Table \ref{table:NR_runs}) increases --- or equivalently, as the orbital angle before merger grows. In contrast, there is no strong correlation between $A$ and the spin of the remnant Kerr BH $(\chi_f)$. Similarly, the time parameter $u_0$ correlates more strongly with $\Omega_{\rm orb}$ than with $\chi_f$.

The power-law exponent $p$ lies in the range $-3.51$ to $-3.79$, which differs from the Price law $u^{-6}$ for $\Psi_4^{\ell=2,m=0}$ \cite{PhysRevD.5.2419,PhysRevD.34.384}.
We propose two possible origins of the tail: 
\begin{itemize}
    \item Intermediate regime of a linear tail. As shown in Fig.~4 of \cite{DeAmicis:2024not}, the tail component requires $u > 10^4$ to fully converge to the Price law. At $u\gtrsim 100$, the exponent ranges from $-3.5$ to $-4.2$ (our exponent for $\Psi_4$ differs from the strain exponent in \cite{DeAmicis:2024not} by 2). The extracted exponents fall within this range and are consistent with this intermediate linear regime.
    \item QNM-driven ``tail''. As discussed in \cite{Cardoso:2024jme,Okuzumi:2008ej}, an outgoing QNM generates a quadratic source that falls polynomially with distance. At second order, it yields a power-law decay in GWs, expected to follow $u^{-4}$ in $\Psi_4$ [Eq.~(66) in \cite{Cardoso:2024jme}\footnote{The exact exponent may remain uncertain, as the authors found discrepancies between some numerical simulations and analytical predictions.}]. Since this decays slower than the Price law, the nonlinearity likely dominates at late times. The presence of quadratic QNMs supports this possibility. 
\end{itemize}
Both channels produce similar power laws in our window\footnote{QNMs also produce power-law decays via other channels  \cite{Green:2013zba,Yang:2013uba}.}.  Distinguishing them requires further study. First, as in \cite{DeAmicis:2024eoy}, one needs to perform a systematic comparison between linear-theory predictions and CCM results across a range of BBH configurations. Second, the QNM-driven channel might establish a
determined link between the tail and the quadratic QNM, similar to
the quadratic-to-linear ratio in
\cite{Khera:2024yrk,Ma:2024qcv}. Using the source term in the second-order Teukolsky equation [see Eq.~(54) in \cite{Ma:2024qcv}], one can make theoretical predictions for the parameters in Eq.~\eqref{eq:power_law_fit} and compare them against our extracted values. We leave these discussions for future work.

%==========================================================================
\section{Conclusion}
\label{sec:conclusion}
We have addressed that CCM, a framework originally proposed in the 1990s \cite{PhysRevLett.76.4303,winicour2012characteristic}, is not only feasible but also enables stable, convergent, and high-accuracy simulations of BBH mergers on an effectively infinite computational domain. By evolving nine BBH systems with the algorithm detailed in Paper I, we showed that their waveforms are in excellent agreement with those from reference simulations with distant outer boundaries, which avoid contamination from inaccurate boundary conditions. In contrast, CCE results exhibited systematic errors. 

Thanks to its unprecedented waveform accuracy, CCM makes it possible to systematically study late-time tails in the fully nonlinear regime. Using rational filters \cite{Ma:2022wpv,Ma:2023cwe,Ma:2023vvr}, we isolated tail components in (quasi-)head-on binary systems from QNMs. Within selected fitting windows, the filtered waveforms were well described by power laws. The tail amplitudes decrease as the initial orbital angular momentum increases, while the power-law exponents range from $-3.51$ to $-3.79$. We proposed two possible origins for the
phenomenon: either the intermediate regime of the linear tail or the QNM-driven nonlinear
tail \cite{Cardoso:2024jme, Okuzumi:2008ej}. To distinguish between these two channels, further work is needed: (i) systematic comparisons between linear calculations \cite{Albanesi:2023bgi,DeAmicis:2024not,Islam:2024vro} and CCM results; and (ii) precise theoretical predictions for the nonlinear tail from second-order black hole perturbation theory \cite{Khera:2024yrk,Ma:2024qcv}. If confirmed as nonlinear, this
would be an example where nonlinearity prevails over linearity at late times.

In this work, we have focused on the tail and ringdown regimes of the Weyl scalar $\Psi_4$ to provide preliminary evidence that CCM effectively removes boundary-induced systematics. A more comprehensive investigation of the GW strain across the full inspiral–merger–ringdown regime is still needed. In particular, it would be interesting to explore whether CCM can resolve tail effects during the inspiral phase through comparisons with post-Newtonian predictions.

Although we have demonstrated that our CCM simulations are stable, convergent, and highly accurate --- at least for the nine BBH systems --- a common concern about CCM in the literature is that it might be only weakly hyperbolic \cite{winicour2012characteristic,Giannakopoulos:2023nrb,Giannakopoulos:2020dih,Giannakopoulos:2021pnh,Giannakopoulos:2023zzm,Gundlach:2024xmo}, which appears to be in tension with our results. Gundlach \cite{Gundlach:2024xmo} has recently provided a tentative solution: by introducing an alternative energy norm, the linearized characteristic initial-boundary value problem about Schwarzschild is shown to be well-posed, and it is conjectured that this result generalizes to arbitrary backgrounds. Given our numerical progress here and in Paper I --- especially its implications for tail and high-accuracy waveform modeling --- it seems timely to address the tension, particularly in the context of the specific CCM formulation we employ.

Finally, as discussed in Paper I and summarized in Sec.~\ref{sec:summary}, the detailed implementation of CCM depends on the underlying Cauchy formalism. For instance, unlike the Cauchy-perturbative matching strategies in \cite{BinaryBlackHoleGrandChallengeAlliance:1997aaw,Rupright:1998uw,Rezzolla:1998xx} or \cite{Szilagyi:2002kv}, the GH formalism \cite{Lindblom:2005qh,Kidder:2004rw} offers a neat decomposition of the forty incoming characteristic fields into three subsets, allowing us to focus only on the ``physical'' subset with just two degrees of freedom. It would be interesting to investigate whether matching this minimal set of dynamical variables is essential to ensure CCM stability, and how CCM can be adapted to other Cauchy formulations.

%==========================================================================
\begin{acknowledgments}
S. M. would like to thank Luis Lehner, Conner Dailey, and Eric Poisson for useful discussions.
 Research at Perimeter Institute is supported in part by the Government of Canada through the Department of Innovation,
Science and Economic Development and by the Province
of Ontario through the Ministry of Colleges and Universities.
This material is based upon work supported by the National Science Foundation
under Grants No.~PHY-2407742, No.~PHY-2207342, and No.~OAC-2209655 at
Cornell. Any opinions, findings, and conclusions or recommendations expressed in
this material are those of the author(s) and do not necessarily reflect the
views of the National Science Foundation. This work was supported by the Sherman
Fairchild Foundation at Cornell.
This work was supported in part by the Sherman Fairchild Foundation and by NSF
Grants No.~PHY-2309211, No.~PHY-2309231, and No.~OAC-2209656 at Caltech.
The computations presented here were conducted in the Resnick High Performance Computing Center, a facility supported by Resnick Sustainability Institute at the California Institute of Technology.
\end{acknowledgments}  

%==========================================================================
\appendix

\section{The characteristic initial data}
\label{app:characteristic_id}
Currently, \texttt{SpECTRE} supports various options for constructing characteristic initial data empirically. Although, in principle, the initial data should be \emph{uniquely} determined from an \textit{ab initio} approach, the default method known as \texttt{ConformalFactor} \cite{ConformalFactor} in \texttt{SpECTRE} has proven to be a suitable choice \cite{Mitman:2024uss}. In particular, it can help reduce initial junk radiation in the memory modes of BBH systems (see discussions below).  
Providing a comprehensive discussion on how characteristic initial data affects
GW waveforms \cite{Bishop:2011iu} and developing an \textit{ab initio}
initial-data solver is beyond the scope of this paper. Here we
 give a brief summary of the \texttt{ConformalFactor}
method, as it was not
fully explained in~\cite{Moxon:2021gbv}.

The \texttt{SpECTRE} characteristic system evolves two scalars and a vector, including 
\begin{itemize}
    \item Bondi $J$: This complex scalar is a volume variable that depends on both angular and radial coordinates. Its evolution is governed by a hierarchical system, see e.g., Eqs.~(14)$-$(18) in \cite{Barkett:2019uae}.
    \item Cauchy angular coordinates $x^A=(\theta,\phi)$: These coordinates form a real vector defined on an extraction worldtube, usually chosen to be a sphere. Cauchy's angular coordinates do not align with those of the characteristic system --- they evolve over time with respect to the characteristic coordinates. The evolution equation can be found in Eq.~(4.53)  of \cite{Ma:2023qjn}. This time-dependent mapping between the two coordinate systems is crucial for variable interpolation on the worldtube. 
    \item Bondi time $\mathring{u}$: The \texttt{SpECTRE} characteristic system adopts the so-called partially flat Bondi-like coordinates, see Table I in \cite{Moxon:2020gha} and Fig.~1 in \cite{Ma:2023qjn}, whose time coordinate differs from the true Bondi time. The two time coordinates are related via $\mathring{u}=\int e^{2\beta}du+\mathring{u}^{(R)}$, see Eq.~(35) in \cite{Moxon:2020gha} for more details. Here $\beta$ is the variable that appears in the Bondi metric, see e.g. Eq.~(1) in \cite{Moxon:2020gha}. The mapping can be interpreted as a BMS transformation: the term involving $e^{2\beta}$ yields time dilation via the conformal factor $\omega$ in Eq.~(2.12a) of \cite{Flanagan:2015pxa}, while $\mathring{u}^{(R)}$ contributes to a supertranslation. Waveform quantities are asymptotically transformed into the true Bondi frame only at the output stage. The BMS transformations for these waveform quantities can be found in Eq.~(94) of \cite{Moxon:2020gha}, consistent with Eq.~(17) of \cite{Boyle:2015nqa}.
\end{itemize}
Initial data are needed for all of these evolved variables. 

For the Bondi time $\mathring{u}$, its initial data sets the angular-dependent integration constant for $\mathring{u}=\int e^{2\beta}du$, which reflects the supertranslation freedom of the first characteristic slice. Currently, the constant is hard-coded to zero in \texttt{SpECTRE}.

The initialization of $J$ is based on the cubic ansatz, see Eq.~(16) of \cite{Moxon:2021gbv}
\begin{align}
    J=\frac{A}{r}+\frac{B}{r^3}. \label{eq:inverse_cubic}
\end{align}
The coefficients $A$ and $B$ are determined by the worldtube data for $J$ and $\partial_rJ$ taken from a Cauchy evolution. Here the term $1/r^2$ is omitted to avoid logarithmic terms, see Sec.~V~B of \cite{Moxon:2020gha}.

The initial data for the Cauchy angular coordinates $x^A$ deserves more attention. Denoting the characteristic angular coordinates as $\hat{x}^{\hat{A}}$, the mapping between them is controlled by the Jacobian $\partial_{\hat{A}}x^A$. In \texttt{SpECTRE}, the Jacobian is represented by two complex scalars $\hat{a}= \hat{q}^{\hat{A}}\partial_{\hat{A}}x^A q_A$ and $\hat{b}=\hat{\bar{q}}^{\hat{A}}\partial_{\hat{A}}x^A q_A$ [see Eq.~(4.13) of \cite{Ma:2023qjn}]
\begin{align}
\partial_{\hat{A}}x^A=
\frac{1}{4}
    \begin{pmatrix}
    \hat{q}_{\hat{A}},\bar{\hat{q}}_{\hat{A}}
    \end{pmatrix}
    \begin{pmatrix}
    \bar{\hat{a}} & \bar{\hat{b}} \\
    \hat{b} & \hat{a}
    \end{pmatrix}
    \begin{pmatrix}
    q^{A} \\
    \bar{q}^{A}
    \end{pmatrix},
\end{align}
where the dyad $q^A$ is given by $(-1,-i\csc\theta)$. Treating the Jacobian as a $2\times2$ matrix, its determinant $\hat{\omega}$ can be computed via 
\begin{align}
\label{eq:conformal_factor}
    \hat{\omega}^2= \frac{1}{2}\epsilon_{AB}\epsilon^{\hat{A}\hat{B}}\partial_{\hat{A}}x^A \partial_{\hat{B}}x^B=\frac{1}{4}(\hat{b}\bar{\hat{b}}-\hat{a}\bar{\hat{a}}),
\end{align}
where $\epsilon_{AB}=\frac{i}{2} q_A\wedge \bar{q}_B$ is the volume form compatible with the unit sphere metric. To obtain Eq.~\eqref{eq:conformal_factor}, we have used $q^A\epsilon_{AB}=-iq_B$.

Similar to  BMS symmetries, where a conformal isometry of the 2-sphere leads to time dilation, see Eq.~(2.12a) of \cite{Flanagan:2015pxa}, the angular diffeomorphism in the present case also leads to the transformation of the Bondi $\beta$ [Eq.~(33a) of \cite{Moxon:2020gha}]
\begin{align}
    e^{2 \hat \beta} = e^{2 \beta} / \hat{\omega},
\end{align}
where the value of $\beta$ on a worldtube comes from a Cauchy evolution, and the transformed $\hat{\beta}$ is defined in the partially flat Bondi-like coordinates, used for future hypersurface integration. \texttt{SpECTRE} now sets the initial value of $\hat{\beta}$ to zero, which yields an algebraic equation for the Jacobian:
\begin{align}
    e^{2\beta}=\hat{\omega}=\frac{1}{2}\sqrt{\hat{b}\bar{\hat{b}}-\hat{a}\bar{\hat{a}}}.\label{eq:initial_condition_omega_hat}
\end{align}
where we have used Eq.~\eqref{eq:conformal_factor}. Since the determinant $\hat{\omega}$ plays a similar role as the conformal factor in BMS, this method is thus termed \texttt{ConformalFactor}. In the code, Eq.~\eqref{eq:initial_condition_omega_hat} is 
iteratively inverted to solve for the scalars $\hat{a}$ and $\hat{b}$, thereby constructing $x^A(\hat{x}^{\hat{A}})$.
Roughly speaking, Eq.~\eqref{eq:initial_condition_omega_hat} sets the initial ``lapse'' at the worldtube to unity, aligning the clock rate with that of inertial observers at future null infinity.

The initial condition for $x^A(\hat{x}^{\hat{A}})$  is crucial in reducing the junk radiation in memory modes, see the red curve in Fig.~10 of \cite{Mitman:2020pbt}. There, a less motivated initial condition $\hat{x}^{\hat{A}}=\delta_A^{\hat{A}} x^A$ was used\footnote{The Bondi $J$ was still constructed using Eq.~\eqref{eq:inverse_cubic}. }.

%%%%%%%%%%%%%%%%%%%%%%%%%%%%%%%%%%%%%%%%%%%%%%%%%%%%%%%%%%%%%%%%%%%%%%%%%%%%%%%
\def\bibsection{\section*{References}}
%%%%%%%%%%%%%%%%%%%%%%%%%%%%%%%%%%%%%%%%%%%%%%%%%%%%%%%%%%%%%%%%%%%%%%%%%%%%%%%
\bibliography{References}

\end{document}